\documentclass[onecolumn,a4paper,accepted=2025-10-30]{quantumarticle}
\pdfoutput=1
\newcommand\rewrite{ \rightarrow }

\newcommand\euler[1]{e^{2 i \pi #1 }}
\newcommand\var[1]{#1}
\newcommand\bool[1]{#1}
\newcommand\rool[1]{#1}
\newcommand\pfcase[2]{\medskip\noindent\textbf{case #1.} #2.\\ \indent}

\newcommand\boolring[1][\mathcal{V}]{\mathbb{F}_{2}[#1]/I_{#1}}

\newcommand\Var{\ensuremath{\text{Var}}}
\newcommand\PS{\ensuremath{\textbf{PS}}}
\newcommand{\GATESET} { \left\{H, X, \text{SWAP}, Z, CZ, \cdots, C^{(m)}Z\right\} }
\newcommand{\OGATESET} { \left\{X, \text{SWAP}, Z, CZ, \cdots, C^{(m)}Z\right\} }
\newcommand{\sem}[1]{\llbracket #1 \rrbracket}
\newcommand{\Null}{ \ensuremath{0}}
\newcommand\HSCITE{\cite{bravyi_simulation_2019,kissinger_simulating_2022,kissinger_classical_2022,pashayan_fast_2022,codsi_cutting_2022,koch_speedy_2023,peres_quantum_2023}}
\newcommand\HOMSTATE[3]{ \ket{ #1 #3 } \bra{ #2 #3 } }

\usepackage{fullpage}
\usepackage{quantikz}
\usepackage{galois}

        \usepackage{amsmath}
        \usepackage{amssymb}
        \usepackage{hyperref}
        \usepackage{amsfonts}
        \usepackage{bbm}
        \usepackage{braket}
        \usepackage{algorithm}
        \usepackage{algpseudocode}
        \usepackage{amsthm}
        \usepackage{tikz}
        \usepackage{tikz-cd}
        \usepackage{stmaryrd}
        \usepackage{xspace,color}
        \usepackage{enumitem}
        \usepackage{scalerel}
        \usepackage{quantikz}
        \usepackage{ifthen}
        \usepackage{currfile}
        \usepackage{cite}
	\usepackage[normalem]{ulem}

        \newtheorem{definition}{Definition}
        \newtheorem{theorem}{Theorem}
        \newtheorem{corollary}[theorem]{Corollary}
        \newtheorem{lemma}[theorem]{Lemma}
        \newtheorem{proposition}[theorem]{Proposition}
        
        \newtheorem{example}[theorem]{Example}
        
        \newtheorem{remark}[theorem]{Remark}

        \newcommand{\unlhdneq}{\mathrel{\ooalign{$\lneq$\cr\raise.22ex\hbox{$\lhd$}\cr}}}
        \newcommand{\rhdneq}{\mathrel{\ooalign{$\gneq$\cr\raise.22ex\hbox{$\rhd$}\cr}}}

        \newcommand{\suchthat}{\, \mid \,}

        \let\P\undefined
        \DeclareMathOperator{\P}{P}
        \DeclareMathOperator{\GI}{GI}

        \DeclareMathOperator{\BQP}{BQP}

        \DeclareMathOperator{\poly}{poly}
        \DeclareMathOperator{\Ima}{Im}

        \newcommand\moduleopen[1][\jobname]{
            \ifdefined \MAIN \else
                \edef\MAIN{\currfilename}
                \title{#1}
                \begin{document}
                \maketitle
            \fi
        }
        \newcommand\moduleclose{
            \ifthenelse{\equal{\MAIN}{\currfilename}}{
                \end{document}
            }{}
        }

    \newcommand\REFS{
        \COMMENT[i]{}{REFS}{orange}
    }
    \newcommand{\NOTE}[2][i]{
        \COMMENT[#1]{#2}{NOTE:}{blue}
    }
    \newcommand{\TODO}[2][i]{
        \COMMENT[#1]{#2}{TODO:}{red}
    }
    \newcommand{\CHECK}[2][m]{
        \COMMENT[#1]{#2}{CHECK:}{purple}
    }
    \newcommand{\COMMENT}[4][m]{
        \if#1m\marginpar{\scriptsize \fi
        \begin{color}{#4}
            \textbf{#3} #2
        \end{color}
        \if#1m } \fi
    }

    \newboolean{INCLUDE_EXTRA}
    \setboolean{INCLUDE_EXTRA}{false}
    \newcommand{\EXTRA}[1]{
        \ifthenelse{\boolean{INCLUDE_EXTRA}}{
            \COMMENT[i]{#1}{EXTRA:}{darkgray}
        }{}
    }

\usepackage{cleveref}
\renewcommand{\vec}[1]{#1}
\definecolor{color1}{HTML}{E66100}
\definecolor{color2}{HTML}{5D3A9B}

\edef\MAIN{\currfilename}

\title{Polynomial-Time Classical Simulation of Hidden Shift Circuits via Confluent Rewriting of Symbolic Sums}

\author{Matthew Amy}
\author{Lucas Shigeru Stinchcombe}
\affiliation{School of Computing Science, Simon Fraser University, Canada}
\begin{document}

\maketitle

\ifdefined \MAIN \else
    \documentclass{article}
    
    \moduleopen
\fi
\begin{abstract}
Implementations of Roetteler's shifted bent function algorithm have
in recent years been used to test and benchmark both classical simulation algorithms
and quantum hardware.
These circuits have many favorable properties, including a tunable amount of non-Clifford
resources and a deterministic output, and moreover do not belong to any class of
quantum circuits that is known to be efficiently simulable.
We show that this family of circuits can in fact be simulated in
polynomial time via symbolic path integrals.
We do so by endowing symbolic sums with a confluent rewriting system
and show that this rewriting system suffices to reduce the circuit's path integral to the hidden shift in polynomial time.
We hence resolve an open conjecture about the efficient simulability of this class of circuits.

\end{abstract}

\moduleclose

\section{Introduction}\label{sec:intro}
\ifdefined \MAIN \else
    \documentclass{article}
    
    \moduleopen
\fi

The classical simulation of quantum circuits is an important problem, both for practical issues of testing and benchmarking quantum circuits, devices, and algorithms, as well as for quantifying and elucidating the nature of quantum speedup. One of the most celebrated results in the classical simulation of quantum circuits showed that circuits comprised of only Clifford gates are polynomial-time simulable. On the other hand, given circuits over a universal gate set such as the well-known Clifford+$T$ basis, an efficient, generic classical simulation algorithm would imply $\mathbf{BQP}= \mathbf{P}$, carrying with it enormous impacts on both classical and quantum computation. 

In a recent series of work \cite{bravyi_trading_2016,bravyi_improved_2016,bravyi_simulation_2019}, it was shown that given a small number of non-Clifford resources, classical simulation of a quantum circuit can be performed in polynomial time. In this sense, non-Clifford resources can be seen as drivers of quantum advantage. The key observation was that gate teleportation reduces the simulation of a Clifford+$T$ circuit to the simulation of a Clifford circuit with a non-Clifford resource state proportional in size to the number of non-Clifford gates. This resource state can then be decomposed as a sum of $k$ stabilizer states and simulation can then proceed by performing $k$ independent, polynomial-time Clifford simulations. The decomposition of resource states into sums of stabilizer states has since been extensively studied \cite{qassim_improved_2021,peleg_lower_2022,kocia_classical_2021,lovitz_new_2022}, but the best-known upper bounds remain exponential in the number of non-Clifford gates to be implemented by teleportation.

Despite its exponential scaling in the number of non-Clifford resources, stabilizer decompositions have shown favorable performance in practical simulations of large circuits. In \cite{bravyi_improved_2016}, stabilizer decomposition-based simulation was performed on a class of simulation benchmarks derived from an oracle algorithm due to Roetteler \cite{rotteler_quantum_2010}. Roetteler's algorithm, which computes a bitstring $s$ hidden in a pair of Maiorana-McFarland bent functions provided as oracles to the quantum computer, was used as a benchmark for two reasons; one that it is a deterministic algorithm, and secondly that it allowed for the precise control over the number of non-Clifford gates since the non-oracle part of the circuit is strictly Clifford. To control the number of non-Clifford resources, instances of the algorithm were generated using degree $3$ constructions of Maiorana-McFarland bent functions, which correspond directly to implementations over the $\{Z, CZ, CCZ\}$, and hence Clifford+$T$, gate set. It can be noted that such instances \emph{require} implementation over a universal gate set, and hence do not fall under known classically simulable classes like Clifford \cite{aaronson_improved_2004}, CNOT-dihedral \cite{cross_scalable_2016}, or commuting circuits \cite{ni_commuting_2013}.

Following its use in \cite{bravyi_improved_2016}, this class of implementations of Roetteler's algorithm has been used extensively to benchmark classical simulation methods \cite{amy_towards_2018,bravyi_simulation_2019,kissinger_simulating_2022,kissinger_classical_2022,pashayan_fast_2022,codsi_cutting_2022,koch_speedy_2023,peres_quantum_2023}.
Through these works however, empirical
evidence that it is not a good choice of benchmark in general began to surface.
In \cite{amy_towards_2018}, instances of the benchmark were simulated deterministically in a fraction of the time by rewriting the circuit's {(symbolic) path integral} or \emph{path sum}.
Benchmark parameters which took hours to simulate probabilistically using stabilizer decompositions were reduced to the hidden shift in seconds via rewriting. Likewise, in \cite{koch_speedy_2023}, which used tensor contraction methods in
the closely related
ZX-calculus \cite{coecke_interacting_2008},
observed that it would contract completely without
any need for stabilizer decomposition. A year earlier the same effect was
observed in \cite{codsi_cutting_2022}, and was conjectured to be polynomial
time simulable.

In this paper, we show that the family of hidden shift circuits from \cite{bravyi_improved_2016} is indeed polynomial-time simulable. Our methods are based on static rewriting of the circuit's path sum. We show that by restricting the rewriting rules of \cite{amy_towards_2018} to control the degree of polynomials of the circuit path sum gives a \emph{confluent} rewrite system which reduces the path sum of the hidden shift circuit family to the shift in provably polynomial-time. Combining simplification of the path sum with explicit evaluation yields a complete simulation method, thereby answering the conjecture by Codsi in \cite{codsi_cutting_2022} in the affirmative. We further show that simulation remains polynomial-time for a larger class of circuit instances based on Maiorana-McFarland bent functions of bounded degree. Conversely, our simulation methods fail to be tractable for functions of unbounded degree, or for implementations of the oracle which are exponentially smaller than the algebraic (polynomial) normal form of the given bent functions.

The rest of the paper is as follows: \Cref{sec:prelim} introduces Roetteler's shifted bent function algorithm, along with the class of concrete circuit implementations we are interested in. In \Cref{sec:pathsums}, we introduce path sums, a rewrite system, and accompanying simulation algorithms. In \Cref{sec:confluence}, we show that the rewrite system is confluent, and in \Cref{sec:simulation} we give sufficient conditions for when the simulation algorithms are efficient, and show that hidden shift circuits satisfy this condition. Finally, \Cref{sec:conclusion} concludes the paper.

\moduleclose

\section{Hidden Shift Circuits and the Shifted Bent Function Problem}\label{sec:prelim}
\ifdefined \MAIN \else
    \documentclass{article}
    
    \moduleopen
\fi

We begin with an overview of Roetteler's algorithm for the shifted bent function problem, 
devised in \cite{rotteler_quantum_2010}
as a problem that gives oracle separations between the complexity classes $\BQP$ and $\P$.
Briefly, the problem involves an oracle which gives access to two Boolean functions $f,g$ such that
$g(x) = f(x + s)$. This oracle is said to ``hide'' the bitstring $s$,
and the problem is to compute the hidden bitstring by querying $f$ and $g$. 
It was shown that for bent functions taken from a family of Boolean functions known as the \emph{Maiorana-McFarland} family,  
exponentially many classical queries but only linearly many quantum queries are required to compute the shift.
Roetteler further showed that if the oracle also gives access to the \emph{dual} of $f$,
a weaker separation of $O(n)$ classical to $O(1)$ quantum queries is obtained, this time
by a \emph{deterministic} quantum algorithm. Standard methods \cite{hallgren_superpolynomial_2008}
then suffice to extend this to a super-polynomial separation.
The hidden shift circuit family used to benchmark simulation algorithms
is an implementation of the deterministic hidden shift algorithm, where
the oracle is ``de-oraclized'' by giving a
concrete circuit implementation. We will briefly introduce the
algorithm as presented in \cite{rotteler_quantum_2010},
and give a construction of the benchmark which captures
and generalizes those used in \HSCITE.

We first recall some theory from the Fourier analysis of Boolean functions.
We will use both Boolean groups $(\mathbb{F}_{2} , +)$
and $(\{1, -1\}  , \cdot )$ as convenient. They are isomorphic by the mappings
\[
    (\mathbb{F}_{2}, +)\galois{(-1)^{(\cdot)}}{\frac{1 - (\cdot)}{2}}(\{1, -1\}, \cdot)
\]
Recall also that the set of pseudo-Boolean functions
$\{f: \mathbb{F}_{2}^{n} \rightarrow \mathbb{R}\}$ forms a $2^n$-dimensional real vector space
under the usual notion of function addition.
We can endow the vector space with an inner product defined as follows
\begin{align*}
    \langle f, g \rangle = \mathbb{E}[fg]  = \frac{1}{2^{n}}\sum_{x \in \mathbb{F}_{2}^{n}} f(x)g(x)
\end{align*}
Observe that the functions $\{\chi_{s} \suchthat s \in \mathbb{F}_{2}^{n}\}$
where $\chi_{s}(x) = (-1)^{s \cdot x}$ form a basis of the pseudo-Boolean functions,
which we call the parity basis.
Indeed it can be seen to be a basis since they are orthonormal and $2^{n}$-many.
The Fourier transform of a pseudo-Boolean function $f$ is then defined
as
\begin{align*}
    \widehat{f} (s) = \langle f, \chi_{s} \rangle
\end{align*}
so that $f$ is represented over $\{\chi_{s}\}$ by
\begin{align*}
    f(x) = \sum_{s \in \mathbb{F}_{2}^{n}} \widehat{f}(s)\chi_{s}
\end{align*}
and the values $\widehat{f}(s)$ are called the Fourier coefficients of $f$.
Now let $f: \mathbb{F}_{2}^{n}\rightarrow \{1, -1\}$ be a Boolean function.
The function $f$ is called a \textit{bent} function if its Fourier coefficients
have the same magnitude.
\begin{definition}\label{def:bent}
    Let $f: \mathbb{F}_{2}^{n}\rightarrow \{1, -1\}$
    be a Boolean function. Then $f$ is bent if for all $s \in \mathbb{F}_{2}^{n}$,
    $\lvert \widehat{f}(s) \rvert = \frac{1}{\sqrt{2^n}}$.
\end{definition}
\Cref{def:bent} essentially states that the Fourier transform $\widehat{f}$ is also a Boolean
function up to a scaling.
For a bent function $f$, its \textit{dual} is a Boolean function defined as
$\overline{f}(x) = \sqrt{2^{n}}\widehat{f}(x)$.
Note that the dual of $\overline{f}$ is again $f$.
The results of \cite{rotteler_quantum_2010} pertain to
a construction of bent functions called the Maiorana-McFarland bent
functions whose definition we reproduce here.

\begin{proposition}[Maiorana-McFarland Bent function] \label{MMfunc}
    Let \(f: \mathbb{F}_{2}^{2n} \rightarrow \mathbb{F}_{2}\)
    be a Boolean function such that
    \begin{align*}
        f(x, y) = &\langle x, \pi (y) \rangle + f_0(y) \\
        \text{where } &f_0: \mathbb{F}_{2}^{n} \rightarrow \mathbb{F}_{2}
        \text{ and } \pi \in \mathcal{S}_{2^{n}} \text{ is a permutation.}
    \end{align*}
    Then $f$ is bent and has the dual function
    \(\overline{f}(x,y) = \langle \pi^{-1}(x), y \rangle + f_0(\pi^{-1}(x))\).
    We call $f$ a Maiorana-McFarland bent function, and the
    set of all such functions $\mathcal{M}$.
\end{proposition}
\begin{proof}
    By direct computation of the Fourier coefficient
    \begin{align*}
        \widehat{f}(s,t)
        &= \frac{1}{2^{2n}}\sum_{x,y \in \mathbb{F}_{2}^{n}}
        (-1)^{\langle x, \pi(y)  + s \rangle + f_0(y) + \langle y, t \rangle } \\
        &= \frac{1}{2^{2n}}\sum_{x,y \in \mathbb{F}_{2}^{n}}
        (-1)^{\langle x, y  + s \rangle + f_0(\pi^{-1}(y))
        + \langle \pi^{-1}(y), t \rangle } \\
        &= \frac{1}{2^{n}}
        (-1)^{f_0(\pi^{-1}(s)) + \langle \pi^{-1}(s), t \rangle } \\
        &= \frac{1}{2^{n}}
        (-1)^{ \langle s, \pi(t) \rangle + f_0(\pi^{-1}(s)) }
    \end{align*}
    where the step from line 2 to line 3 follows from the fact that
    the summands where $y \neq s$, occur with both positive and negative
    factor $2^{n-1}$ times each and hence cancel, while the summands
    where $y = s$ occur with positive factor exactly $2^{n}$ times.
    Thus we have
    \(\lvert \widehat{f}(s,t) \rvert = \frac{1}{2^n}\), and its dual
    is as desired.
\end{proof}

Note that the permutation $\pi \in S_{2^{n}}$ is any bijection
$\pi: \mathbb{F}_{2}^{n} \rightarrow \mathbb{F}_{2}^{n}$
which will be restricted when we discuss the hidden shift circuit
family when considered in the context of benchmarking classical simulations.
Now we can formally define the shifted bent function problem, which is the task
of computing a shift $s \in \mathbb{F}_{2}^{n}$ hidden by a bent function.

\begin{definition}[Shifted Bent Function Problem]
    For a given \(s \in \mathbb{F}_{2}^{n}\), and a bent function
    \(f: \mathbb{F}_{2}^{n} \rightarrow \mathbb{F}_{2}\), set $g: x \mapsto f(x +s)$.
    With oracle access to $f, g$, compute the shift
    \(s\) where oracles are implemented by following unitaries
    \[
        O_{G}: \ket{x} \mapsto (-1)^{g(x)} \ket{x} ,
        \; \; \; O_{F}: \ket{x} \mapsto (-1)^{\overline{f}(x)} \ket{x} 
    \]
The phase oracles above can equivalently be constructed from oracles
computing in the computational basis with the usual method of phase kickback 
via the state \(\ket{-} = \frac{\ket{0} - \ket{1} }{\sqrt{2}}\).
\end{definition}

Roetteler \cite{rotteler_quantum_2010} showed that for any bent function $f$,
the shifted bent function problem as defined above can be solved deterministically
with two oracle calls on a quantum computer. On the other hand, in the classical case,
Roetteler also showed that for a Maiorana-McFarland bent function $f$,
classically $\Theta(n)$ queries are necessary and sufficient to compute the shift.
The quantum algorithm solving the shifted bent function problem is given by
the following circuit.
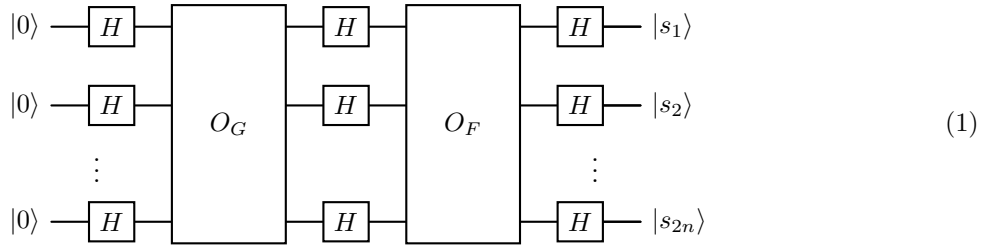
\begin{equation}\label{roetteler_algorithm}
    \begin{quantikz}[row sep=0.5cm]
        \lstick{$\ket{0}$} & \gate{H}
        & \gate[4]{O_{G}} \hphantom{wide} & \gate{H}
                & \gate[4]{O_F} \hphantom{wide} & \gate{H} & \qw \rstick{$\ket{s_{1}}$}\\
        \lstick{$\ket{0}$} & \gate{H}
        &  & \gate{H}
                &  & \gate{H} & \qw \rstick{$\ket{s_{2}}$}\\
        & \lstick{$\vdots$} \wireoverride{n} & \wireoverride{n} & \wireoverride{n} & \wireoverride{n} & \rstick{$\vdots$} \wireoverride{n} \\
        \lstick{$\ket{0}$} & \gate{H}
        &  & \gate{H}
                &  & \gate{H} & \qw \rstick{$\ket{s_{2n}}$}
    \end{quantikz}
\end{equation}

In ~\cite{bravyi_simulation_2019}, Bravyi and Gosset benchmarked their
simulation algorithm on instances of the quantum circuit
~\ref{roetteler_algorithm} for randomly generated
Maiorana-McFarland bent functions where the permutation
$\pi \in S_{2^{n}}$ was fixed to be the identity. In addition,
the bent functions were of degree at most $3$
where the \emph{degree}
is the polynomial degree of the Boolean function expressed as a multilinear
polynomial.
The oracles were implemented over the gate set
$\{X, Z, CZ, CCZ\}$, which were then compiled to
Clifford+T circuits via
$CCZ = (I \otimes I \otimes H) \text{Toff} (I \otimes I \otimes H)$
and a gadget implementing the Toffoli gate $\text{Toff}$ over Clifford+T with $4$ $T$ gates due to 
Jones~\cite{jones_low-overhead_2013}. The choice to use the measurement-based implementation of 
\cite{jones_low-overhead_2013} was motivated by the lower $T$-count than direct implementations, 
as their simulation algorithm scales exponentially in the number of $T$ gates.
The same construction of the circuit has since been used to benchmark a variety of simulation methods \HSCITE, as well as benchmarking quantum hardware
~\cite{lubinski_application-oriented_2023,wright_benchmarking_2019}.
More generally, we can define instances of the hidden shift circuit
for functions of degree $m+1$ by extending the gate set to include multiply-controlled $Z$ gates as well as allowing permutations $\pi \in S_{n}$,
which is to say bijections
$\pi: \mathbb{F}_{2}^{n} \rightarrow \mathbb{F}_{2}^{n}$ achieved by
permuting bits. Note for such permutations, we have the following
equality of the inner product function
$\langle \cdot, \cdot \rangle: \mathbb{F}_{2}^{2n} \rightarrow \mathbb{F}_{2}$.
\begin{align*}
    \langle \pi^{-1}(x), y \rangle =
    \langle x, \pi(y) \rangle
\end{align*}

\begin{definition}[Hidden Shift Circuit Family] \label{bravyi-gosset-roetteler circuits}
    Let $\pi \in \mathcal{S}_n$ be a permutation,
    $f_0: \mathbb{F}_{2}^{n} \rightarrow \mathbb{F}_{2}$ be a Boolean function,
    and $ s = s_1s_2\in \mathbb{F}_{2}^{2n}$ be a bitstring.
    A hidden shift circuit $C_{(\pi, f_0, s)}$ for the triple
    $(\pi, f_0, s)$ is any circuit of the form \ref{roetteler_algorithm}
    on $2n$ qubits where oracles
    \begin{align*}
        &O_G: \ket{x,y} \mapsto (-1)^{\langle x+ s_1, \pi(y+s_2) \rangle + f_0(y + s_2)}
        \ket{x,y} \\
        &O_F: \ket{x,y} \mapsto (-1)^{\langle x, \pi(y) \rangle + f_0(\pi^{-1}(x))}
        \ket{x,y}
    \end{align*}
    are implemented over the gate set $\OGATESET$. In particular this implies that
    $f_0$ has degree at most $m+1$.
\end{definition}
The fact that $m$ is a fixed positive integer will be essential
in the proof of polynomial-time classical simulation as polynomials
of bounded degree have only polynomially-many terms in the number of variables
when expressed in their multilinear form.
Previous instances of the circuit construction \HSCITE were restricted to the 
Bravyi-Gosset parameters
$m=2$ and $\pi = \mathrm{id}$.
Note that $C^{(m)}Z$ is non-Clifford whenever $m\geq 2$, and in particular requires more non-Clifford resources as $m$ increases. When $\mathrm{deg}(f_0)\geq 3$, Roetteler's algorithm moreover \emph{requires} a non-linear Boolean operation, i.e. a $CCZ$ or $CCX$ gate, in order to implement the oracles. As $\{H, CCX\}$ is universal for quantum computing \cite{aharonov_simple_2003}, \emph{any} circuit implementation must be over a universal gate set. The circuit family defined above thus constitutes a family of non-trivial circuits which (1) generates an exponential-size superposition of basis states, (2) generates an intermediate state which is highly-entangled, (3) uses interference to produce the correct result, and (4) cannot be implemented over a classically-simulable set of quantum gates. These features are known \cite{nielsen_quantum_2011} to be necessary (but not sufficient) conditions for quantum speedup, as together they preclude polynomial-time classical simulation by standard methods.

Correctness of Roetteler's algorithm can be shown by direct calculation, which corresponds to a sequence of simplifications due to interference yielding the final shift~$s$. While simpler proofs are possible, the calculation below is effectively automatable as we show in following sections.
Writing $g$ for the $s$-shifted $f$ which is computed by $O_G$, $\overline{f}$ again for the dual of $f$ computed by $O_F$, and letting $U_{(\pi, f_0, s)}$ be the unitary implemented by circuit $C_{(\pi, f_0, s)}$ we have
\begin{align*}
    U_{(\pi, f_0, s)}\ket{0}
    &= \frac{1}{2^{3n}}\sum_{x,y,z\in\mathbb{F}_2^{2n}}(-1)^{g(x) + \langle x,y\rangle + \overline{f}(y) + \langle y, z\rangle}\ket{z} \\
    &= \frac{1}{2^{3n}}\sum(-1)^{\langle x_1+ s_1, \pi(x_2+s_2) \rangle + f_0(x_2 + s_2) + \langle x,y\rangle + \overline{f}(y) + \langle y, z\rangle}\ket{z} \\
    &= \frac{1}{2^{3n}}\sum(-1)^{\langle x_1, \pi(x_2+s_2) + y_1 \rangle + \langle  s_1, \pi(x_2+s_2)\rangle + f_0(x_2 + s_2) + \langle x_2,y_2\rangle + \overline{f}(y) + \langle y, z\rangle}\ket{z} \\
    &= \frac{1}{2^{2n}}\sum(-1)^{\langle s_1, \pi(x_2+s_2)\rangle + f_0(x_2 + s_2) + \langle x_2,y_2\rangle + \overline{f}(\pi(x_2+s_2),y_2) + \langle \pi(x_2+s_2), z_1\rangle + \langle y_2, z_2\rangle}\ket{z}
\end{align*}
where the last equality follows because $\sum_{x_1,y_1\in\mathbb{F}_2^n} (-1)^{\langle x_1, \pi(x_2+s_2) + y_1 \rangle} = 2^n$ if $y_1 = \pi(x_2+s_2)$, and $0$ otherwise, and we leave the variables summed over implicit. Expanding $\overline{f}(\pi(x_2+s_2),y_2)$ we get $\langle x_2+s_2, y_2\rangle + f_0(x_2+s_2)$ and in particular the two $f_0(x_2+s_2)$ terms cancel. The remaining calculation follows below:
\begin{align*}
    U_{(\pi, f_0, s)}\ket{0} 
    &= \frac{1}{2^{2n}}\sum(-1)^{\langle s_1, \pi(x_2+s_2)\rangle + \langle x_2,y_2\rangle + \langle x_2+s_2, y_2\rangle + \langle \pi(x_2+s_2), z_1\rangle + \langle y_2, z_2\rangle}\ket{z} \\
    &= \frac{1}{2^{2n}}\sum(-1)^{\langle s_1 + z_1, \pi(x_2+s_2)\rangle + \langle y_2, z_2 + s_2\rangle}\ket{z} \\
    &=\ket{s_1,s_2}
\end{align*}
where again the last equality follows by noting that $\sum_{y_1,z_2\in\mathbb{F}_2^n} (-1)^{\langle y_2, z_2 + s_2 \rangle}$ destructively interferes whenever $z_2\neq s_2$, and likewise $\sum_{x_2,z_1\in\mathbb{F}_2^n} (-1)^{\langle s_1 + z_1, \pi(x_2+s_2) \rangle}$ destructively interferes when $z_1\neq s_1$.

It is worth noting that for the specific constructions of hidden shift circuits used in ~\cite{bravyi_improved_2016}, there is a similar polynomial length proof of correctness by circuit equalities, which we give in Appendix~\ref{appendix_hiddenshift_circuit_proof}.
The circuit equality proof relies not only on the structure of the construction, but also on a particular sequence of circuit rewrites which does not in general correspond to an effective simulation --- or even circuit simplification --- method. Instead, we codify the above proof using path sums, where we find that the algebraic equalities employed in the above proof correspond to a \emph{confluent} rewrite system, in that every sequence of rewrites results in the same normal form. Thus the above proof will witness that a simulation algorithm which arbitrarily simplifies the codified expression using these algebraic equalities is guaranteed to yield the hidden shift, including for more general cases where the hidden shift circuit is not implemented literally as constructed in ~\cite{bravyi_improved_2016}.

\moduleclose

\section{Path Sums} \label{sec:pathsums}
\ifdefined \MAIN \else
    \documentclass{article}
    
    \moduleopen
\fi

We now turn our attention to path sums, which form the basis of our simulation methods.
Path sums, corresponding to a discretization of Feynman's path integral,
are symbolic representations of linear operators as finite sums
of parameterized linear operators over variables ranging in $\mathbb{F}_{2}$.
The analysis of circuits via discretized path integrals, colloquially termed the 
\emph{sum-over-paths} technique, has seen application over the years in a great deal
of contexts, notably proving classical complexity results
\cite{dawson_quantum_2004,montanaro_quantum_2017}.
In \cite{amy_towards_2018} the sum-over-paths technique was formalized as a mathematical
object --- called a \emph{path sum} --- and equipped with a rewriting system which allowed it
to be statically reduced without explicitly expanding the sum. The path sum has since been extended
in various ways~\cite{amy_towards_2018,amy_symbolic_2022,amy_complete_2023,
vilmart_structure_2021,vilmart_completeness_2023}; these
various incarnations as rewrite systems have been shown to be complete
for the Clifford \cite{vilmart_structure_2021} and Toffoli-Hadamard \cite{vilmart_completeness_2023}
fragments, as well as general $\mathcal{R}$-linear operators with \emph{unbalanced} amplitudes \cite{amy_complete_2023}. Beyond completeness results,
path sums parameterized
on classical data such as natural numbers,
called \emph{higher order path sums},
were introduced in ~\cite{chareton_automated_2021}
to verify families of circuits by deductive program verification.

Path sums are represented concretely as collections of polynomials giving the
phase, inputs, and outputs along a \emph{path} --- a particular assignment to the variables
over which the sum is defined.
We restrict our attention to polynomials over
$\mathbb{F}_{2}$, corresponding to operators over the universal Toffoli-Hadamard
gate set, as this simplification has the benefit that all polynomials
are of the same polynomial ring over $\mathbb{F}_{2}$.
This fragment was previously studied and denoted
\textbf{SOP}$[\frac{1}{2}] $ in ~\cite{vilmart_completeness_2023},
where a rewrite system was given which was \emph{complete} when viewed
as an equational theory, but not \textit{confluent}.

\begin{definition}[Boolean Path Sum]
    A (Boolean) path sum is an expression of the form
    \[
        s \sum_{V}(-1)^{P} \ket{O_1, \ldots , O_m} \bra{I_1, \ldots, I_n}
    \]
    where

    \begin{enumerate}
        \item \(s \in \mathbb{C}\)  is a complex scalar
        \item \(V\)  is a set of indeterminates over $\mathbb{F}_{2}$
        \item \(P, O_{1}, \dots, O_{m}, I_{1}, \dots, I_{n}\)
            are multivariate polynomials in $\boolring[V]$
    \end{enumerate}
where $I_{V}$ is the ideal generated by $v^{2} - v$ for all $v \in V$.
\end{definition}
Recall that taking the quotient by the ideal generated by $v^{2} -v$
expresses the fact that $v^{2}$ (and hence all positive powers of $v$)
are identified with $v$.
While we use the notation $s \sum_{V}(-1)^{P} \ket{O_1, \ldots , O_m} \bra{I_1, \ldots, I_n}$ to express the intuition of a path sum as a linear operator, a path sum is equally
well-presented by the tuple
\[
    (V, s, P, O:=(O_1,\dots, O_m), I:=(I_1,\dots, I_n)).
\]

The set of all path sums we will call $\PS$, and
polynomials which are elements of $\mathbb{F}_{2}[V]/I_{V}$
we will refer to as \textit{Boolean polynomials}. Note that Boolean polynomials
may be represented uniquely as \emph{multilinear polynomials} ---
that is polynomials which are degree $\leq 1$ in any given variable.
For an expression $A \in \PS$, we will use the convention that
the various components of the expression are labeled as follows.
\[
    A = s_{A} \sum_{V_{A}} (-1)^{P_{A}} \ket{O_A} \bra{I_A}
\]
Furthermore, we will call \(A: \left| I_{A} \right| \rightarrow \left| O_{A} \right| \),
the signature of $A$. All path sums will have an interpretation
as a linear operator in the following way by computing the sum
over the variables $V_{A}$ as they range in $\mathbb{F}_{2}$.
In particular for $\lvert V_A\rvert = k$, $\lvert I_{A} \rvert ~=n$,
and $\lvert O_{A} \rvert  = m$
we may view $P_{A}, I_{A}, O_{A}$
as functions on $\vec{v} \in \mathbb{F}_{2}^{k}$ where
\begin{align*}
    P_{A}&: \mathbb{F}_{2}^{k} \rightarrow \mathbb{F}_{2} \\
    I_{A}&: \mathbb{F}_{2}^{k} \rightarrow \mathbb{F}_{2}^{n} \\
    O_{A}&: \mathbb{F}_{2}^{k} \rightarrow \mathbb{F}_{2}^{m}
\end{align*}
respectively,
the corresponding linear operator
is $eval(A): \mathbb{C}^{2^n} \rightarrow \mathbb{C}^{2^m}$
defined~by

\begin{align}~\label{eval}
    eval(A) = s_{A} \sum_{\vec{v} \in \mathbb{F}_{2}^{k}}
        (-1)^{P_{A}(\vec{v})} \ket{O_A(\vec{v})} \bra{I_A(\vec{v})}
\end{align}

As path sums correspond to symbolic representations of linear operators,
we can define composition and the tensor product of path sum expressions
in the following way.

\begin{definition}[Composition, Tensor product, Identity, Zero, Adjoint]\label{composition}
  Without loss of generality we assume in the following that for any $A,B\in \PS$ that $\Var(A) \cap \Var(B) = \emptyset$, and that $y_i\notin \Var(A) \cup \Var(B)$ for any $i$.
  We define
    \begin{enumerate}
        \item for \(A: m \rightarrow n\), \(B: k \rightarrow m \in \PS\),
            composition \(A \circ B : k \rightarrow n\)
            \begin{align*}
                A \circ B = \frac{s_{A}s_{B}}{2^m}
                    \sum_{V_A\cup V_B \cup \left\{ y_1, \ldots, y_m \right\} }
                    (-1)^{P_{A} + P_{B} +
                    \sum_{i=1}^{m} y_i\left( O_{B,i} + I_{A,i} \right)}
                    \ket{O_{A}} \bra{I_{B}}
            \end{align*}
        \item for \(A: m \rightarrow n, B: k \rightarrow l \in \PS\) the
            tensor product \(A \otimes B: m+k \rightarrow n+l\)
            \begin{align*}
                A \otimes B = s_{A}s_{B}
                    \sum_{V_A\cup V_B }
                    (-1)^{P_{A}+ P_{B}}
                    \ket{O_{A}O_{B}} \bra{I_{A}I_{B}}
            \end{align*}
        \item for each \(n \in \mathbb{Z}_{\geq 0}\) the identity, \(I_n\)
            \begin{align*}
                I_n = \sum_{\left\{ y_1, \ldots, y_n \right\} }
                (-1)^{0}
                \ket{y_1,\ldots,y_n} \bra{y_1,\ldots,y_n}
            \end{align*}
        \item for each \(m,n \in \mathbb{Z}_{\geq 0}\) the zero, \(0_{m,n}\) with signature
            $m \rightarrow n$.
            \begin{align*}
                \Null_{m,n} = 0\sum_{\emptyset} (-1)^{0}
                \ket{\underbrace{0,\ldots,0}_{n}} \bra{\underbrace{0,\ldots,0}_{m}}
            \end{align*}
        \item for \(A: m \rightarrow n \in \PS\) the
            adjoint \(A^{\dagger} : n \rightarrow m\)
            \begin{align*}
                A^{\dagger} = s_{A}^{\ast}
                    \sum_{V_A}
                    (-1)^{P_{A}}
                    \ket{I_{A}} \bra{O_{A}}
            \end{align*}
    \end{enumerate}
\end{definition}

The composition, tensor products, and adjoint on path sums
can be seen to be sound with respect to $eval$, in the sense that
for arbitrary $A, B \in \PS$
\begin{align*}
    eval(A \otimes B) &= eval(A) \otimes eval(B) \\
    eval(A \circ B) &= eval(A) \circ eval(B) \\
    eval(A^{\dagger}) &= eval(A)^{\dagger}
\end{align*}
whenever they are well-defined with respect to their signatures.
Indeed, for composition notice that the term $y_{i}(O_{B,i} + I_{A,i})$
asserts that $O_{B,i} = I_{A,i}$ for all $i$, since whenever $O_{B,i} + I_{A,i} = 1$,
summing over $y_{i}$ destructively interferes, and constructively interferes
when $O_{B,i} + I_{A,i} = 0$.
Note also that a sum over the empty set corresponds to the absence of a sum, i.e.
\begin{align*}
    eval(s \sum_{\emptyset} (-1)^{c} \ket{d_{1}, \dots, d_{n}} \bra{b_{1}, \dots, b_{m}}) = s(-1)^{c} \ket{d_{1}, \dots, d_{n}} \bra{b_{1}, \dots, b_{m}}
\end{align*}
where $c, d_{i}, b_{j} \in \mathbb{F}_{2}$.
Having equipped path sums with identities and parallel and sequential composition, we
may interpret circuits $C\in (\mathcal{G}, I, \otimes, \cdot)$
over a gate set $\mathcal{G}$ by giving an interpretation of the gates in $\mathcal{G}$.
Given a mapping $\sem{\cdot}:\mathcal{G}\rightarrow \PS$
such that for $g \in \mathcal{G}$ with signatures $g:m\rightarrow n$
we say that $\sem{\cdot}$ is well-formed
if $\sem{g}:m\rightarrow n$, i.e. the signatures of $\sem{g}$ match. We extend
well-formed interpretations $\sem{\cdot}$ to circuits
$C\in (\mathcal{G}, I, \otimes, \cdot)$ in the obvious way:
\begin{align*}
    \sem{I_n} &= I_n \\
    \sem{g} &= \sem{g} \\
    \sem{g_2\cdot g_1} &= \sem{g_2}\circ \sem{g_1} \\
    \sem{g_1\otimes g_2} &= \sem{g_1}\otimes \sem{g_2}
\end{align*}

Given linear operators $U_g:\mathbb{C}^{2^m}\rightarrow \mathbb{C}^{2^n}$
for each $g:m\rightarrow n\in\mathcal{G}$,
if $eval({\sem{g}}) = U_g$ for every $g\in\mathcal{G}$, it follows that $eval({\sem{C}}) = U_C$
for any circuit $C$ over the gate set $\mathcal{G}$ --- that is, $\sem{C}$ is a sound interpretation of $C$. We define a sound interpretation of $\mathcal{G}= \GATESET$ in $\PS$ as follows
\begin{align*}
    \llbracket H  \rrbracket &=
    \frac{1}{\sqrt{2}} \sum_{\left\{ x, y \right\} }
        (-1)^{xy}\ket{y} \bra{x}\\
    \llbracket X  \rrbracket &=
    \sum_{\left\{ x \right\} }
        \ket{x+1} \bra{x}\\
    \llbracket C^{(m)}Z  \rrbracket &= 
    \sum_{\left\{ x_1, \ldots x_m,y \right\} }^{}
        (-1)^{y \prod_{i=1}^{m} x_i}
        \ket{ x_1, \ldots x_m,y} \bra{ x_1, \ldots x_m,y} \\
    \llbracket SWAP \rrbracket  &= \sum_{\left\{ x, y \right\} }
        \ket{y, x} \bra{x, y}
\end{align*}

\begin{example}[]
    The following path sum interpretations for two Hadamard
    gates in sequence and in parallel
    illustrate the path sum composition and tensor rules
    respectively and can be seen to evaluate to their
    standard matrix interpretations.
    \begingroup
    \addtolength{\jot}{0.5em}
    \begin{align*}
        \llbracket
            \begin{quantikz}[row sep=0.5cm]
                \qw & \gate{H} & \gate{H} & \qw
            \end{quantikz}
        \rrbracket \;
        & = \left(
            \frac{1}{\sqrt{2}} \sum_{\{x_{1}, z_{1}\} }
                (-1)^{x_{1}z_{1}}
                \ket{x_{1}} \bra{z_{1}}
        \right)
        \circ \left(
            \frac{1}{\sqrt{2}} \sum_{\{x_{2}, z_{2}\} }
                (-1)^{x_{2}z_{2}}
                \ket{x_{2}} \bra{z_{2}}
        \right)
        \\
        &= \frac{1}{4}\sum_{\{x_{1},x_{2},z_{1},z_{2},y\} }
                (-1)^{x_{1}z_{1} + x_{2}z_{2} + y(z_{1} + x_{2})}
                \ket{x_{1}} \bra{z_{2}}
        \\
        &\xmapsto{eval}
        \;\begin{bmatrix}
            1 & 0 \\
            0 & 1
        \end{bmatrix}
        \\
        \Biggl\llbracket
            \begin{quantikz}[row sep=0.3cm]
                \qw & \gate{H} & \qw \\
                \qw & \gate{H} & \qw
            \end{quantikz}
        \Biggr\rrbracket \;
        & = \left(
            \frac{1}{\sqrt{2}} \sum_{\{x_{1}, z_{1}\} }
                (-1)^{x_{1}z_{1}}
                \ket{x_{1}} \bra{z_{1}}
        \right)
        \otimes \left(
            \frac{1}{\sqrt{2}} \sum_{\{x_{2}, z_{2}\} }
                (-1)^{x_{2}z_{2}}
                \ket{x_{2}} \bra{z_{2}}
        \right)
        \\
        & = \frac{1}{2} \sum_{\{x_{1}, x_{2}, z_{1}, z_{2} \} }
                (-1)^{x_{1}z_{1} + x_{2}z_{2}}
                \ket{x_{1}, x_{2}} \bra{z_{1}, z_{2}}
        \\
        &\xmapsto{eval}
        \;\frac{1}{2} \begin{bmatrix*}[r]
            1   & 1     & 1     & 1     \\
            1   & -1    & 1     & -1    \\
            1   & 1     & -1    & -1    \\
            1   & -1    & -1    & 1
        \end{bmatrix*}
        \\
    \end{align*}
    \endgroup
\end{example}

For any circuit $C$ with $n$ qubit lines
and $k$ gates, we will denote its \emph{volume} by $\lvert C \rvert$
which is the product of the number of qubit lines and number of gates,
\emph{i.e.} $\lvert C \rvert = nk$. Note that sometimes the depth $d$
of a circuit is used in place of the number of gates to define volume.
Since $d \leq k \leq kn$, these notions are polynomially related.

\begin{proposition}\label{prop:circuit_to_expression}
    Let $\mathcal{G}=\GATESET$ for some fixed $m$.
    For any circuit $C$ over $\mathcal{G}$, $\llbracket C \rrbracket $ can be computed in time
    and has size polynomial in the volume $\lvert C \rvert$ of $C$. Furthermore,
    $\llbracket C \rrbracket $ has at most $O(\lvert C \rvert)$ variables.
\end{proposition}
\begin{proof}
    Let $C$ have $n$ qubit lines and $k$ gates
    and notice that the size of any path sum expression
    $A \in \PS$ over $\mathcal{G}$
    is at most polynomial in $\lvert V_{A} \rvert $. Indeed,
    the Boolean polynomials of $A$ have degree at most $m+1$.
    Then the number of terms of a polynomial in $\mathbb{F}_{2}[V_{A}]/I_{V_{A}}$
    of degree $\leq m+1$ is bounded by
    \begin{align*}
        \binom{\lvert V_{A} \rvert }{0}
        + \binom{\lvert V_{A} \rvert }{1}
        + \dots 
        + \binom{\lvert V_{A} \rvert }{m+1}
        = O(\lvert V_{A} \rvert^{m+1} )
    \end{align*}
    which bounds the size of $A$ by $O(n \lvert V_{A} \rvert^{m+1})$.
    Each composition $C = A \circ B$ has number of variables
    $\lvert V_{C} \rvert  = \lvert V_{A} \rvert + \lvert V_{B} \rvert  + O(n)$
    while tensor products preserve the number of variables.
    Thus at the end of the procedure we have applied at most $k$
    compositions introducing at most $O(kn)$ new variables.
    Since each $g \in \mathcal{G}$ has at most $m+1$  variables,
    we have variable count of the final expression of $\leq O(kn)+ (m+1)k = O(kn)$
    giving a final size of
    \begin{align*}
        O(n (kn)^{m+1})
    \end{align*}
    which takes time $k \cdot O(n(kn)^{m+1}) = O((kn)^{m+2})$ for computing
    $\leq k$ intermediate expressions.
\end{proof}

\subsection{Affine equivalence of path sums}
A simple observation of path sum evaluation ~\ref{eval} reveals that
for an invertible affine map
$M: \mathbb{F}_{2}^{k} \rightarrow \mathbb{F}_{2}^{k}$, the evaluation of
$A$ is preserved by a transformation $M \vec{v}$ on the points
$\vec{v} \in \mathbb{F}_{2}^{k}$
\begin{align}~\label{pathsum_evaluation}
    eval(A) = s_{A} \sum_{\vec{v} \in \mathbb{F}_{2}^{k}}
        (-1)^{P_{A}(M\vec{v})} \ket{O_A(M\vec{v})} \bra{I_A(M\vec{v})}
\end{align}
since the sum is commutative and $M$ is invertible.
However, we can also choose to view
the transformation of points instead as a transformation on the
polynomials of our path sum $A$, which is to say a transformation of
path sum $A$ yielding path sum $B$ with components such that
\begin{align*}
    P_{B}(\vec{v}) &= P_{A}(M\vec{v}) \\
    O_B(\vec{v}) &= O_A(M\vec{v}) \\
    I_B(\vec{v}) &= I_A(M\vec{v})
\end{align*}
When $M \vec{v}= L \vec{v} + \vec{b}$ is the representation
of $M$ where $L$ is a linear map and $\vec{b} \in \mathbb{F}_{2}^{k}$
is a translation, this is achieved
by an isomorphism on the polynomials of $A$
which sends $v_{i} \mapsto L_{i} \vec{v} + b_{i}$
where $L_{i}, b_{i}$ are the $i$-th rows of $L, \vec{b}$ respectively.

Viewed in this way, path sums which only differ by an affine translation
of the evaluation points $\vec{v} \in \mathbb{F}_{2}^{k}$
can be captured by a ring isomorphism applied uniformly on all constituent
polynomials of the expression.
The observation shows that we need only consider \textit{degree-preserving}
isomorphisms $\phi$ which is to say the total degree $\deg(\phi(v)) = 1$ for all $v \in V$.
In order to be precise about what degree-preserving means in the context of Boolean
rings $\mathbb{F}_{2}[V]/I_{V}$ in which the polynomials of our path sums reside,
we will make use of a functor which will allow us to carry over these notions
which are defined unambiguously in $\mathbb{F}_{2}[V]$.

For a homomorphism $\phi: \mathbb{F}_{2}[V] \rightarrow \mathbb{F}_{2}[W]$
which is \textit{degree-nonincreasing}, that is,
the total degree $\deg(\phi(v)) \leq 1$ for all indeterminates $v \in V$,
$\phi$ induces a unique homomorphism
$\phi' : \mathbb{F}_{2}[V]/I_{V} \rightarrow \mathbb{F}_{2}[W]/I_{W}$
as described in the following diagram which preserves composition, identities
and inverses.
\begin{equation}~\label{multilinear_morphism_diagram}
    \begin{tikzcd}[wire types={n,n}, row sep=huge, column sep=huge, scale=2]
    \mathbb{F}_{2}[V] \arrow[r, "\phi" ] \arrow[d, "\pi "']
    & \mathbb{F}_{2}[W] \arrow[d, "\pi"] \\
    \mathbb{F}_{2}[V]/I_{V} \arrow[r, "\phi'"', dashed]
    & \mathbb{F}_{2}[W]/I_{W}
\end{tikzcd}
 \end{equation}
In an abuse of notation we will refer to both
$\phi$ and the uniquely induced $\phi'$ as simply $\phi$.
With these technical considerations out of the way, we can now
formally define affine transformations of path sums.

\begin{definition}[Affine Transformation of Path Sums]
    Let $A \in \PS$ be an arbitrary expression,
    and $\phi:\mathbb{F}_{2}[V_{A}] \rightarrow \mathbb{F}_{2}[W] $
    be a degree-preserving ring isomorphism
    such that $\lvert V_{A} \rvert = \lvert W\rvert$.
    We call $\phi$ an \textit{affine transformation} of path sums and
    $\phi (A)$ the affine transformation
    of $A$ by $\phi $, defined by
    \[
        \phi(A) = s_{A} \sum_{W} (-1)^{\phi(P)}
            \ket{\phi(O_A)}  \bra{\phi(I_A)}
    \]
    where $\phi(O_A) = (\phi(O_{A,1}), \dots, \phi(O_{A,m}))$ and likewise for $\phi(I_A)$.
\end{definition}

In this work, we are concerned with path sums where
all relevant polynomials are over $\mathbb{F}_{2}$. However,
affine transformations can be extended to more elaborate path sums,
by a Boolean lifting described in ~\cite{amy_towards_2018}.
For instance, for the Clifford path sums in ~\cite{vilmart_structure_2021},
where path sums had the form
\begin{align}~\label{clifford_path_sum}
    s \sum_{V} \euler{
        \frac{1}{8}\bool{P^{(0)}} + \frac{1}{4}\bool{P^{(1)}} + \frac{1}{2}\bool{P^{(2)}}
    } \ket{O} \bra{I}.
\end{align}
where $\bool{P^{(k)}} \in \mathbb{F}_{2}[V]$ is a Boolean polynomial limited to degree at most $k$,
we see that an affine transformation preserves the form ~\ref{clifford_path_sum}.
As shown for linear substitutions in~\cite{amy_towards_2018},
this is a consequence of the fact that more generally for fixed $m$,
affine transformations preserve expressions of the form
\begin{align*}
    s \sum_{V} \euler{(
        \frac{1}{2^{m}}\bool{P^{(0)}} + \cdots  + \frac{1}{2^k}\bool{P^{(m-k)}} 
        + \cdots  + \frac{1}{2}\bool{P^{(m-1)}})
    } \ket{O} \bra{I}.
\end{align*}
And thus a bound on the time complexity of evaluating the path sum
based on the degrees of the polynomials involved holds for the entire equivalence
class of expressions that differ only by an affine transformation.

It will often be necessary
to refer to the variables of a polynomial which we define formally as follows.

\begin{definition}[]~\label{polynomial_variables}
    Let $K$ be a field and $f \in K[\mathcal{V}]$.
    Then $\Var(f)$ is defined as the minimal subset
    $V \subseteq \mathcal{V}$ such that $f \in K[V]$.
    If $f_1,\ldots, f_n$ are polynomials
    (in potentially different polynomial rings), we define
    \begin{align*}
        Var(f_1,\ldots, f_n) = \bigcup_{i = 1  }^{n}Var(f_i).
    \end{align*}
\end{definition}

Although many of our lemmas will be stated in general for affine transformations,
the main result of this work will be with respect to a simpler notion
of equivalence which will correspond to when $L$ of $M \vec{v} = L \vec{v} + \vec{b}$
is a permutation matrix. This can be thought of as renaming variables, potentially
with an affine translation by a scalar, and can be restated in terms
of polynomial isomorphisms.
Whenever a degree-nonincreasing homomorphism
$\phi : \mathbb{F}_{2}[V] \rightarrow \mathbb{F}_{2}[W]$
is such that $\lvert \Var(\phi(v)) \rvert \leq 1$, for all indeterminates
$v \in V$, we say that $\phi$ is \textit{simple}.
\begin{definition}[Simple Transformation of Path Sums]
    Let $A \in \PS$ be an arbitrary expression,
    and $\phi:\mathbb{F}_{2}[V_{A}] \rightarrow \mathbb{F}_{2}[W] $
    be a simple degree-preserving isomorphism
    such that $\lvert V_{A} \rvert = \lvert W\rvert$.
    Then we call $\phi$ a simple transformation and $\phi(A)$
    the simple transformation of $A$ by $\phi$.
\end{definition}

We will make frequent use of the fact that
for a simple transformation $\phi(A) = B$,
for any $x \in V_{A}$, the restriction of the domain of $\phi$
to $\mathbb{F}_{2}[V_{A} \setminus \{x\}]$
has image contained in $\mathbb{F}_{2}[V_{B} \setminus \{z\}]$
for some $z \in V_{B}$. Simple transformations
naturally yield the equivalence relation $\sim$ which we will use
throughout this work.

\begin{definition}[Simple Equivalence]\label{affine-equivalence}
   For $A, B \in \PS$, we say that $A \sim B$ if and only if
     there exists a simple transformation $\phi$
    such that $\phi(A)~=~B$.
\end{definition}

\begin{example}[]~\label{example_hh_simple_equivalence}
    Consider the path sum interpretation of the quantum
    circuit consisting of two Hadamard gates in sequence.
    The simple transformation $\phi$ defined by
    \begin{align*}
        \phi:
        u \mapsto u +1, \quad
        w \mapsto w, \quad
        x \mapsto z + 1, \quad
        y \mapsto y + 1, \quad
        z \mapsto x + 1,
    \end{align*}
    witnesses the following simple equivalence
    \begin{align*}
        \llbracket
            \begin{quantikz}[row sep=0.5cm]
                \qw & \gate{H} & \gate{H} & \qw 
            \end{quantikz}
        \rrbracket
        \;&=
        \frac{1}{4}\sum_{\{u,w,x,y,z\} }
            (-1)^{zw + xy + uw + uy}
            \ket{z} \bra{x} \\
        &\sim
        \frac{1}{4}\sum_{\{u,w,x,y,z\} }
            (-1)^{xw + zy + z + uw + uy + u}
            \ket{x+1} \bra{z+1}
    \end{align*}
    The simple transformation corresponds to the translation
    of evaluation points $M$ as in~\cref{pathsum_evaluation}
    defined by the following with respect to the ordering
    $u,w,x,y,z$.
    \begin{align*}
        M: v \mapsto \begin{bmatrix}
            1 & 0 & 0 & 0 & 0 \\
            0 & 1 & 0 & 0 & 0 \\
            0 & 0 & 0 & 0 & 1 \\
            0 & 0 & 0 & 1 & 0 \\
            0 & 0 & 1 & 0 & 0
        \end{bmatrix}
        v
        + \begin{bmatrix}
            1 \\ 0 \\ 1 \\ 1 \\ 1
        \end{bmatrix}
    \end{align*}
\end{example}

Simple equivalence is comparable to $\alpha$-equivalence
in symbolic logic, with an additional equivalence made between a sum over
a variable $x \in \mathbb{F}_2$ and the sum over its negation $1 + x \in \mathbb{F}_2$,
and in this sense can be thought of as an \emph{internal} equivalence of path sums.
More generally, affine equivalence can be viewed as an equivalence relation
for Boolean path sums which was enforced in~\cite{vilmart_structure_2021}
by the (bra) and (ket) rules. There, by virtue of the restriction to
path sums of $\textbf{SOP}_{\textit{Clif}}$, it was shown that all path sums
can be rewritten by the rewrite system $ \rightarrow_{\textit{Clif+}}$
to a form where all variables occurring in (non-zero) path sums
occur in the $O_{i}, I_{j}$ state polynomials.
Then by the fact that the degrees of polynomials
$O_{i}, I_{j}$ for path sums of $\textbf{SOP}_{\textit{Clif}}$
are always $1$, a canonical choice
of the equivalence class was made by applying affine transformations
allowed by the (bra) and (ket) rules using the ordering
of polynomials
$O_{1} \prec \dots \prec O_{m} \prec I_{1} \prec \dots \prec I_{n}$.

It can be seen that deciding simple equivalence for path sums in general
is $\GI$-hard. Indeed, there is a standard reduction from the graph isomorphism
problem to deciding whether two quadratics are equivalent up to a permutation
of variables.
We can embed this quadratic into a path sum, and with the help of
an additional polynomial,
we can force simple equivalence to be a permutation. Specifically, we can construct
a path sum $S_{G}$ for a graph $G = (V,E)$ with vertices $V = \{v_{1}, \dots, v_{n}\} $
viewed as indeterminates as follows.
\begin{align*}
    S_{G} = \sum_{V} (-1)^{\prod_{i}^{n}v_{i}}
    \ket{
        \sum_{(i,j) \in E \suchthat i < j} v_{i}v_{j}
    } 
\end{align*}
Then for undirected graphs $G, H$ without loops on $n$ vertices,
$S_{G} \sim S_{H}$ if and only if there is a graph isomorphism $G \cong H$.

More generally for deciding affine equivalence, the problem has in fact been
studied under the name, the Isomorphism of Polynomials with One Secret
(IP1S) problem~\cite{faugere_polynomial_2006,perret_fast_2005}
where its \emph{two secret} variant, known simply as the Isomorphism
of Polynomials (IP) problem, has been used
as a basis for a public key encryption scheme called the Hidden Field
Equations (HFE)~\cite{patarin_hidden_1996}.
We describe the IP1S problem in brief. In the following,
let $\mathbb{F}_{q}$ be a finite field where $q$ is a prime power,
and $\mathrm{GA}_{n}(\mathbb{F}_{q})$ the general affine group
on the vector space $\mathbb{F}_{q}^{n}$.
The IP1S problem is to compute, given two lists
of polynomials $a = (a_{1},\dots,a_{m}), \: b = (b_{1}, \dots, b_{m})
\in \mathbb{F}_{q}[X]^{m}$ on $n$ indeterminates,
an affine map
$S \in \mathrm{GA}_{n}(\mathbb{F}_{q})$, such that
\begin{align*}
    (b_{1}(x), \dots, b_{m}(x)) = (a_{1}(Sx), \dots, a_{m}(Sx))
\end{align*}
as functions $\mathbb{F}_{q}^{n} \rightarrow \mathbb{F}_{q}^{m}$,
if such an $S$ exists.
The problem of deciding IP1S is known to be
$\GI$-hard~\cite{patarin_improved_1998}
by a reduction from the graph isomorphism problem
to the IP1S decision problem for $q=2$.
Since deciding affine equivalence of path sums is the IP1S
decision problem for $q = 2$ and the list of path sum polynomials
$(P, O:=(O_1,\dots, O_m), I:=(I_1,\dots, I_n))$,
it follows that deciding affine
equivalence of path sums is also $\GI$-hard in general.

\subsection{Rewrite rules}
Having interpreted a circuit $C$ as a path sum $\llbracket C \rrbracket \in \PS$,
simulation will be done by simplifying $\llbracket C \rrbracket $ by a rewrite system
$\rewrite$ which is formally a binary relation on $\PS$.
If simple equivalence, or more generally affine equivalence is \emph{internal},
we can consider rewrites as \emph{external} equivalence of path sums
due to interference.

Rewrite systems for path sums have been studied in~\cite{amy_complete_2023,amy_symbolic_2022,amy_towards_2018,vilmart_completeness_2023,vilmart_structure_2021}.
In \cite{amy_towards_2018}, a rewrite system which was complete for deciding
equivalence of Clifford unitaries was given,
and was generalized to all Clifford path sums in \cite{vilmart_structure_2021}.
Further, \cite{vilmart_completeness_2023}
gave a complete rewrite system for the Boolean fragment we consider in this paper.
The following rewrite system corresponds to a subset of the Clifford-complete system
of \cite{vilmart_structure_2021} restricted to Boolean path sums.
For a variable $v \in V$ and $f \in \mathbb{F}_{2}[V]$,
we will denote by $[v \leftarrow f]$ the substitution of $v$ with $f$.

\begin{definition}[Rewrite System]\label{rewrite_system}
    Let $\rightarrow$ be a binary relation on $\PS$ which is the union
    of binary relations
    $\rightarrow_{elim}, \rightarrow_{z},\rightarrow_{hh}$ defined below.
    \begin{enumerate}
    \item
        Whenever $x \notin Var(P,O,I)$,
        \begin{align}
        s \sum_{V} (-1)^{\rool{\rool{P}}} \ket{O} \bra{I}
            \longrightarrow_{elim}
            2 s \sum_{ V \setminus \left\{ \var{x}\right\} }
            (-1)^{\rool{P}} \ket{O} \bra{I}
            \label{eq:elim}
            \tag{Elim}
        \end{align}
    \item
        Whenever $z \notin Var(R,O,I)$
        \begin{align}
            s\sum_{V} (-1)^{
                    \var{z} + \rool{R}
                } \ket{O} \bra{I}
                \longrightarrow_{z} 
                0 \sum_{\emptyset } (-1)^{0}
                    \ket{0,\ldots,0} \bra{0,\ldots,0}
                    \label{eq:z}
                    \tag{Z}
        \end{align}
    \item
        Whenever $x \notin Var(R,O,I), y \notin Var(Q)$ and $\lvert \Var(Q) \rvert  \leq 1$
        \begin{align}
            s\sum_{V} (-1)^{\var{x}
                ( \var{y} + \bool{Q} ) +\rool{R}} \ket{O} \bra{I}
                \longrightarrow_{hh}
                \left(s\sum_{V \setminus \{\var{x}\} }
                    (-1)^{R}
                    \ket{O} \bra{I} \right)
                    [\var{y} \leftarrow \bool{Q}]
                    \label{eq:hh}
                    \tag{HH}
        \end{align}
\end{enumerate}
 \end{definition}

We will call the variable $x$ in the precondition for the HH rule,
the \textit{pivot} variable. Note that since polynomials
are assumed to be presented as \emph{multilinear} polynomials (polynomials
with degree at most $1$ in any given variable), we have $x \notin \Var(y + Q)$.
Further, $\Var(Q) \leq 1$ implies that $\deg(Q) \leq 1$ since a multilinear
polynomial in a single variable is just a linear polynomial, and a constant
when there are no variables.
Like in the case of affine transformations of path sums which
were essentially polynomial isomorphisms, we can view the ~\ref{eq:z} rule and ~\ref{eq:hh} rules
as inducing polynomial homomorphisms. Indeed, the~\ref{eq:z} rule induces the zero homomorphism
$\mathbb{F}_{2}[V] \rightarrow \mathbb{F}_{2}$ which sends everything to zero.
Likewise, the ~\ref{eq:hh} rule induces a substitution homomorphism
$\mathbb{F}_{2}[V] \rightarrow \mathbb{F}_{2}[V \setminus \{x\}]$.
While in previous rewrite systems the HH rule had no restriction on the structure of $Q$,
here we restrict $\lvert \Var(Q) \rvert \leq 1$ so that the substitution
$[y \leftarrow Q]$ is simple.

\begin{example}[]~\label{example_hh_rule_simplifying_two_hadamards}
    Consider the path sum interpretation of the quantum
    circuit consisting of two Hadamard gates in sequence
    as in Example~\ref{example_hh_simple_equivalence}.
    The~\ref{eq:hh} rule along with the~\ref{eq:elim} rule
    can be seen to enable the rewrite to the identity
    path sum $I_{1}$. The pivot and eliminated variables
    are marked in red at each rewrite.
    \begin{align*}
        \llbracket
            \begin{quantikz}[row sep=0.5cm]
                \qw & \gate{H} & \gate{H} & \qw 
            \end{quantikz}
        \rrbracket
        \;&=
        \frac{1}{4}\sum_{\{u,w,x,y,z\} }
            (-1)^{zw + xy + \textcolor{red}{u}(w+y)}
            \ket{z} \bra{x} \\
        &\rewrite_{hh}
        \frac{1}{4}\sum_{\{w,x,y,z\} }
            (-1)^{\textcolor{red}{y}(z + x)}
            \ket{z} \bra{x} \\
        &\rewrite_{hh}
        \frac{1}{4}\sum_{\{\textcolor{red}{w},x,z\} }
            (-1)^{0}\ket{x} \bra{x} \\
        &\rewrite_{elim}
        \frac{1}{2}\sum_{\{x,\textcolor{red}{z}\} }
            (-1)^{0}\ket{x} \bra{x} \\
        &\rewrite_{elim}
        \sum_{\{x\} }
            (-1)^{0}\ket{x} \bra{x} \;=\; I_{1}
    \end{align*}
\end{example}

We can think the restriction of $\lvert \Var(Q) \rvert \leq 1$
in the \ref{eq:hh} rule as
going hand-in-hand with the chosen notion of simple equivalence.
That is, if we were to
trade the restriction $\lvert \Var(Q) \rvert \leq 1$ with
 $\deg(Q) \leq 1$, then we would naturally arrive
at \textit{affine} equivalence rather than simple equivalence,
whereby two path sums are equivalent if they are related by an affine transformation.
Simple homomorphisms are desirable as for any polynomial $P$, not only is
the substitution $P[y \leftarrow Q]$ degree-preserving, but also preserves
the number of variables $\Var(P)$. As a method of \emph{simplifying} path sums,
the restriction to simple HH avoids complications where an affine HH may
obstruct subsequent rule applications by introducing variables into
$\Var(R,O,I)$ which are inessential.
The following example illustrates this complication of allowing $\deg(Q) \leq 1$,
as it relates to presentations of polynomials.
The presence of both $z, w$ in the second path sum
can be seen to be inessential, in that there exists an affine transformation
such that the polynomials are presented with fewer variables, for instance the transformation
which sends $z \mapsto z + w$.
\begin{example}[]
    Let $ \rightarrow_{hh}$ have the restriction on $Q$
    that $\deg(Q) \leq 1$. Then
    $A = \frac{1}{4 \sqrt{2}}\sum_{\{w,x,y,z\} } (-1)^{x(y + z + w) +y} \ket{y}$
    rewrites to
    \begin{align*}
        A &\rightarrow_{hh} \frac{1}{4 \sqrt{2}}\sum_{\{w,y,z\} } (-1)^{y} \ket{y} \\
        A &\rightarrow_{hh} \frac{1}{4 \sqrt{2}}\sum_{\{w,y,z\} } (-1)^{z + w} \ket{z + w}
    \end{align*}
    where the results of the rewrites are related by an affine transformation
    but not a simple transformation.  Furthermore, the first path sum admits
    two eliminations
    while the second admits only one.
\end{example}

Since the RHS of the Z rule is identical for all path sums of the same signature,
we will denote it $A \rightarrow_{z} \Null$.
We will let $\overset{\ast}{ \rightarrow }$ be the reflexive-transitive closure
of $ \rightarrow$ throughout.
A property of interest for rewrite systems is whether each rewrite sequence
is of finite length which guarantees the existence of \textit{normal forms}
(not necessarily unique) which are expressions
that do not admit any rewrite.

\begin{definition}[Noetherian]
    A rewrite system $ \rightarrow $ is noetherian
    iff there is no infinite sequence
    $x_{1} \rightarrow \cdots \rightarrow x_{n} \rightarrow \cdots $.
\end{definition}

\begin{lemma}~\label{noetherian}
    The rewrite system of Definition ~\ref{rewrite_system} is noetherian
\end{lemma}
\begin{proof}
    Trivial since each rewrite rule removes at least one variable.
\end{proof}

\begin{corollary}~\label{polynomial_rewrites}
    A rewrite sequence $A_{0} \rewrite A_{1} \rewrite \dots \rewrite A_{l}$
    has length at most polynomial in the number of variables $\lvert V \rvert$ of $A_{0}$.
\end{corollary}
The above actually shows that the rewrite system is \textit{bounded}
according the definitions given in ~\cite{huet_confluent_1980}, which is to say for every
given path sum $A$, all rewrite sequences starting from $A$ are at most some finite length.

\subsection{Simulation}~\label{section_simulation}

At this point, we can already state a general simulation algorithm for circuits $C$
over gate set $\mathcal{G}$. For a circuit $C$ implementing a unitary $U$,
we consider the task of strong simulation --- given
computational basis states $\ket{\vec x}, \ket{\vec y}$, compute the output amplitude
$\bra{\vec y} U \ket{\vec x}$ whose magnitude squared is the probability of observing $\ket{\vec y}$
on input $\ket{\vec x}$.

\begin{algorithm}[H]
    \caption{$\PS$ strong simulation for quantum algorithms}\label{alg:strong_sim}
    \begin{algorithmic}
        \Require $n$-qubit Circuit $C$ over the gate set $\mathcal{G}$ and strings
        $\vec{x},\vec{y} \in \mathbb{F}_{2}^n$
        \State $f \gets \bra{y_1,y_2,\ldots, y_n} \circ
            \llbracket C\ \rrbracket \circ \ket{x_1,x_2,\ldots, x_n}$
        \While{$f \rightarrow f'$}
            \State $f \gets f'$
        \EndWhile
        \State \Return $eval(f)$
    \end{algorithmic}
\end{algorithm}

Hidden shift circuits are deterministic circuits and strong simulation
would in the worst case,
still require $2^n$-many invocations of algorithm~\ref{alg:strong_sim}
for each value of $\vec{y}$ to find the unique shift $s$
that returns amplitude 1. Indeed, each invocation would comprise of the hidden
shift circuit $C$ with input $x = 0 \dots 0$, and a choice of $y$
which returns $1$ if and only if $y$ is exactly the shift $s$.
More generally for quantum algorithms with measurement on a subset of qubits,
strong simulation is also not sufficient to simulate their outputs as in general,
one must compute all of the relevant amplitudes by ~\cref{alg:strong_sim}.
We can modify algorithm ~\ref{alg:strong_sim} to accommodate these scenarios
where measurement occurs on one qubit by the following, and can easily be
extended for subsets of qubits. We note that unlike~\cref{alg:strong_sim},
~\cref{alg:bqp_sim} returns a probability.

\begin{algorithm}[H]
    \caption{$\PS$ 1-qubit measurement simulation for quantum algorithms}\label{alg:bqp_sim}
    \begin{algorithmic}
        \Require $n$-qubit Circuit $C$ over the gate set $\mathcal{G}$, string
        $\vec{x} \in \mathbb{F}_{2}^n$ and measurement index $i \in [n]$
        \State $g \gets \llbracket C\ \rrbracket \circ \ket{x_1,x_2,\ldots, x_n}$
        \State $f \gets
            g^{\dagger}
            \circ \left( I_{i-1} \otimes \ket{1}\bra{1} \otimes I_{n-i} \right)
            \circ g$
        \While{$f \rightarrow f'$}
            \State $f \gets f'$
        \EndWhile
        \State \Return $eval(f)$
    \end{algorithmic}
\end{algorithm}

We note that when post-selection in circuits is allowed by use of projectors,
then~\cref{alg:bqp_sim} is just a strong simulation on a circuit of the form
$C = U^{\dagger} P U$ with projector $P$ and input strings $\vec{x} = \vec{y}$.
We also note that for a uniform family of circuits $\{C_{n}\}$ deciding a language
in $\BQP$, a classical algorithm deciding the same language amounts
to
accepting input $\vec{x} \in \mathbb{F}_{2}^{n}$ iff
~\cref{alg:bqp_sim} outputs a value $\geq 2/3$ on input $C_{n}$ and $\vec{x}$.
We can understand algorithms~\ref{alg:strong_sim} and~\ref{alg:bqp_sim}
to consist of essentially two phases.
The first phase is to symbolically rewrite the path sum until a normal
form is reached, and the second consists of an evaluation.
One can notice by the rewrite system of Definition~\ref{rewrite_system},
that path sums remain polynomially-sized over the entire first phase.
This is to say that for a circuit $C$ over $\mathcal{G}$,
whenever
\begin{align*}
    \sem{C} \overset{\ast }{\rightarrow }A
\end{align*}
where $\overset{\ast }{ \rightarrow}$ is the
reflexive-transitive closure of $ \rightarrow $,
$A$ has size bounded by $O(\lvert C \rvert^{m+2})$. Indeed,
the bound on the size of $\sem{C}$ of
Proposition~\ref{prop:circuit_to_expression}
was due to the degree bound of all of its constituent polynomials.
Since none of the rewrite rules of Definition~\ref{rewrite_system}
increase polynomial degrees,
the same bound holds for $A$. Furthermore, by
Corollary~\ref{polynomial_rewrites}, we have that the while loop executes
at most polynomially-many times, making the first phase terminate necessarily
in polynomial time, since each rewrite takes polynomial time
to identify and apply.
The costly portion is the second evaluation phase which is exponential in
the number of variables, which by
proposition~\ref{prop:circuit_to_expression}
is $\Omega(\lvert C \rvert)$ in the worst case.

Algorithms~\ref{alg:strong_sim} and~\ref{alg:bqp_sim} always succeed,
but in the worst case take exponential time in the volume of the circuit,
due to the evaluation which takes exponential time in the number of variables.
In the subsequent sections, in light of the preceding discussion on the
efficiency of the first simplification phase, we will develop sufficient conditions
for when these algorithms are efficient, which will boil down to
demanding that the first phase sufficiently simplifies the path sum
to reduce the complexity of the costly evaluation phase. Further,
by showing our rewrite system to be \emph{confluent}, it will follow
that any choice of $f \rightarrow f'$ at each iteration of the loop
ultimately yields an equivalent path sum in the evaluation phase,
removing the need for any specific simplification strategy.

\moduleclose

\section{Confluent Rewriting for Path Sums}\label{sec:confluence}
\ifdefined \MAIN \else
    \documentclass{article}
    
    \moduleopen
\fi

Having armed ourselves with a ring theoretic view of path sums,
we are in a position to prove \textit{confluence} of our rewrite system.
The property of confluence as seen
in Figure~\ref{confluence_modulo_equivalence_figure}
can be intuitively understood as the property that ``all roads lead to Rome''.
It is desirable as we don't ever need to backtrack while simplifying our expression
and coupled with the noetherian property of a rewrite system, we are guaranteed
unique normal forms so that our motto then becomes
``all roads lead to Rome in finite time''.
It is readily seen that all of our rewrites can be
identified and applied in polynomial time and since we've seen already that
any rewrite sequence terminates in polynomially-many steps,
this further improves our motto to
\begin{align*}
    \text{``all roads lead to Rome in polynomial time''}
\end{align*}
or more accurately, ``all rewrites lead to a unique normal form in polynomial time''.
The rewrite system of~\cite{vilmart_completeness_2023} was not confluent
and only complete when the rewrite system was made into an equational theory
after taking the symmetric closure.
We will show that the rewrite system ~\ref{rewrite_system}
is confluent modulo simple equivalence.
Confluence modulo an equivalence relation has the formal following definition,
where $\overset{\ast }{ \rightarrow }$ denotes the reflexive-transitive
closure of a binary relation $ \rightarrow $.

\begin{figure}
    \caption{Confluence Modulo $\sim $ of Definition~\ref{confluence_modulo_equivalence}}
        ~\label{confluence_modulo_equivalence_figure}
    \begin{align*}
        \begin{tikzpicture}[>=Stealth, thick]
            \node (X) at (-0.5, 2) {$A$};
            \node (W) at (0.5, 2) {$B$};
            \node (Y) at (-1.5, 0.5) {$C$};
            \node (Z) at (1.5, 0.5) {$D$};
            \node (Yp) at (-0.5, -1) {$C'$};
            \node (Zp) at (0.5, -1) {$D'$};
            \draw[->] (X) -- (Y) node[midway, left] {$\ast$};
            \draw[->] (W) -- (Z) node[midway, right] {$\ast$};
            \draw[->, dashed] (Y) -- (Yp) node[midway, left] {$\ast$};
            \draw[->, dashed] (Z) -- (Zp) node[midway, right] {$\ast$};
            \draw[-,
                decorate,
                decoration={snake, amplitude=0.5mm, segment length=4mm}]
                (Yp) to (Zp);
            \draw[-,
                decorate,
                decoration={snake, amplitude=0.5mm, segment length=4mm}]
                (X) to (W);
        \end{tikzpicture}
     \end{align*}
\end{figure}
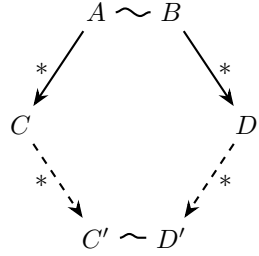

\begin{definition}[Confluence Modulo $\sim$]~\label{confluence_modulo_equivalence}
    If $\sim $ is an equivalence relation, then a rewrite system
    $\rewrite $ is confluent modulo $\sim $ iff
    for all $A, B$ such that $A \sim B$,
    and for any $C, D$  such that
    $A \overset{\ast }{\rewrite} C, B \overset{\ast }{\rewrite} D$,
    there exists $C', D'$ with
    $C' \sim D'$ and
    $C \overset{\ast }{\rewrite} C', D \overset{\ast }{\rewrite} D'$.
\end{definition}

Definition ~\ref{confluence_modulo_equivalence}
is summarized by Figure~\ref{confluence_modulo_equivalence_figure}.
In the case of noetherian relations which we have by Lemma~\ref{noetherian},
we can consider a simpler characterization
called local confluence modulo $\sim $.

\begin{definition}[Local Confluence Modulo $\sim $]~\label{local_confluence_modulo_equivalence}
    A rewrite system $\rewrite$ is locally confluent modulo $\sim $
    iff conditions $\alpha $ and $\beta $ are satisfied:
    \begin{align*}
        \alpha&: \forall ABC \quad A \rewrite B \land A \rewrite C
            \implies B \overset{\sim }{\downarrow} C \\
        \beta&: \forall ABC \quad A \sim B \land A \rewrite C
            \implies B \overset{\sim }{\downarrow} C
    \end{align*}
    where $B \overset{\sim }{\downarrow } C$ denotes the existence
    of $B', C'$ such that
    $B \overset{\ast }{\rightarrow} B'$,
    $C \overset{\ast }{\rightarrow} C'$
    and $B' \sim C'$.
\end{definition}

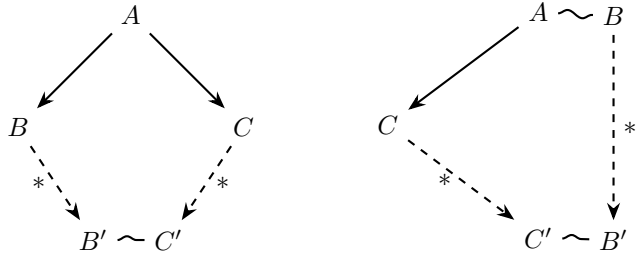
\begin{figure}
    \caption{$\alpha $ and $\beta $ properties of
    Definition~\ref{local_confluence_modulo_equivalence}}
    ~\label{local_confluence_modulo_equivalence_figure}
    \begin{align*}
        \begin{tikzpicture}[>=Stealth, thick]
            \node (X) at (0, 2) {$A$};
            \node (Y) at (-1.5, 0.5) {$B$};
            \node (Z) at (1.5, 0.5) {$C$};
            \node (Yp) at (-0.5, -1) {$B'$};
            \node (Zp) at (0.5, -1) {$C'$};
            \draw[->] (X) -- (Y);
            \draw[->] (X) -- (Z);
            \draw[->, dashed] (Y) -- (Yp) node[midway, left] {$\ast$};
            \draw[->, dashed] (Z) -- (Zp) node[midway, right] {$\ast$};
            \draw[-,
                decorate,
                decoration={snake, amplitude=0.5mm, segment length=4mm}]
                (Yp) to (Zp);
        \end{tikzpicture}
         \quad \quad \quad \quad
        \begin{tikzpicture}[>=Stealth, thick]
            \node (X) at (0, 2) {$A$};
            \node (Y) at (1, 1.95) {$B$};
            \node (Z) at (-2, 0.5) {$C$};
            \node (Zp) at (0, -1) {$C'$};
            \node (Yp) at (1, -1.03) {$B'$};
            \draw[->] (X) -- (Z);
            \draw[->, dashed] (Z) -- (Zp) node[midway, left] {$\ast$};
            \draw[->, dashed] (Y) -- (Yp) node[midway, right] {$\ast$};
            \draw[-,
                decorate,
                decoration={snake, amplitude=0.5mm, segment length=4mm}]
                (Zp) to (Yp);
            \draw[-,
                decorate,
                decoration={snake, amplitude=0.5mm, segment length=4mm}]
                (X) to (Y);
        \end{tikzpicture}
     \end{align*}
\end{figure}

\begin{lemma}[\cite{huet_confluent_1980}, Lemma 2.7]
    Let $\rewrite$ be a noetherian relation. For any equivalence
    $\sim $, $\rewrite$ is confluent modulo $\sim $
    iff $\rewrite$ is locally confluent modulo $\sim $.
\end{lemma}

We will show that the rewrite system satisfies both the $\alpha$ and $\beta$ properties.
The key to proving confluence will be the universal property of the quotient for polynomial rings
--- seemingly different choices of the application of rewrite rules will correspond to
different representations of polynomials modulo a common kernel.
The universal property of the quotient will guarantee that the outcomes are related
by an affine transformation in general, and in the case of the rewrites of
Definition~\ref{rewrite_system}, related by a simple transformation.
We will start by proving the $\beta $ property for cases
$\{ \rightarrow_{elim}, \rightarrow_{z}\}$,
then we'll show how the universal property of the quotient connects to path sums
to prove the $\beta $ property for $ \rightarrow_{hh}$. Finally we'll
borrow further ideas from ring theory to consider what can be said of sequences
of $\rightarrow_{hh}$ and finish by giving a relatively succinct proof
of $\alpha$ property.

\begin{lemma}~\label{beta_elim}
    The ~\ref{eq:elim} rule satisfies property $\beta $
    of Definition~\ref{local_confluence_modulo_equivalence}.
\end{lemma}
\begin{proof}
    Let $x$ be the variable eliminated from $A$ to produce $A'$,
    and $\phi(A) = B$.
    By the definition of simple equivalence it
    follows that the restriction of the domain of $\phi$
    to $\mathbb{F}_{2}[V_{A} \setminus \{x\}]$
    has image in $\mathbb{F}_{2}[V_{B} \setminus \{z\}]$
    for some $z \in V_{B}$, so that $\phi$ with this restriction is
    \begin{align*}
        \phi': \mathbb{F}_{2}[V_{A} \setminus \{x\}]
        \rightarrow \mathbb{F}_{2}[V_{B} \setminus \{z\}]
    \end{align*}
    Thus $z \notin \Var(P_{B}, O_{B}, I_{B})$
    and $z$ can be eliminated from $B$ yielding $B'$.
    It follows that $\phi' $ with exactly
    this restriction is the simple transformation
    such that $\phi'(A') = B'$.
\end{proof}

\begin{lemma}~\label{beta_z}
    The ~\ref{eq:z} rule satisfies property $\beta $
    of Definition~\ref{local_confluence_modulo_equivalence}.
\end{lemma}
\begin{proof}
    Let $\phi$ be a simple transformation of path sums and $A, B \in \PS$
    with $\phi(A) = B$, and let $A$ admit an application of the Z
    rule $A \rewrite A' = \Null$ with
    \begin{align*}
        A  &= s_{A} \sum_{V_{A}}(-1)^{x + R} \ket{O_{A}}\bra{I_{A}}
    \end{align*}
    Then $\phi(x) = z + c$ for some $z \in V_{B}, c \in \mathbb{F}_{2}$,
    and it follows that
    \begin{align*}
        B  &= s_{A} \sum_{V_{A}}(-1)^{ z + c + \phi(R)} \ket{\phi(O_{A})}\bra{\phi(I_{A})}
    \end{align*}
    and $z \notin Var(\phi(R), \phi(O_{A}), \phi(I_{A}))$.
    Thus $B$
    satisfies the precondition for the Z rule, and can be thus be rewritten
    to exactly~$A' = \Null$.
\end{proof}

The $\beta $ property for the HH rule will follow from the fact
that the substitution induced satisfies the universal property for
quotients for the relevant rings.
Recall the definition of the universal property of the quotient
in the category of rings. Let $R$ be a ring and $I \subseteq R$
be an ideal, then the pair $\langle \pi, S \rangle $
where $S$ is a ring, and $\pi: R \rightarrow S$ a ring homomorphism
satisfies the universal property of the quotient for $I$ if
for all $\psi: R \rightarrow T$ such that $\ker \psi \supseteq I$,
there exists a unique $\phi: S \rightarrow T$
such that $\psi = \phi \circ \pi $. For a more in-depth account
of ring theory, the interested reader is encouraged
to consult chapter 3 of ~\cite{aluffi_algebra_2009}.

\begin{lemma}~\label{universal_rewrite}
    Let $V$ be a set of indeterminates containing $x$,
    and let $Q \in \mathbb{F}_{2}[V \setminus \{x\}]$ with
    $\deg(Q) \leq 1$.
    Then the substitution homomorphism
    $\phi: \mathbb{F}_{2}[V] \rightarrow \mathbb{F}_{2}[V \setminus \{x\}]$
    defined by
    \begin{align*}
        \phi: x \mapsto Q
    \end{align*}
    and the identity on $v \in V \setminus \{x\}$,
    has the universal property of the quotient
    $\pi: \mathbb{F}_{2}[V] \rightarrow \mathbb{F}_{2}[V]/(x+Q)$.
    Furthermore, whenever $\phi': \mathbb{F}_{2}[V] \rightarrow \mathbb{F}_{2}[W]$
    is a degree-nonincreasing surjective homomorphism
    with $\ker(\phi') = (x+Q)$, the isomorphism $\psi$ induced
    in the following diagram
    is degree-preserving. If in addition, $\phi'$ is simple
    then $\psi$ is simple.
    \begin{align}~\label{universal_rewrite_diagram}
        \begin{tikzcd}[wire types={n,n}, column sep=2.5cm, row sep=huge,
            ampersand replacement=\&]
            \mathbb{F}_{2}[V]
                \arrow[r, "{\phi}"]
                \arrow[dr, "\phi'"] \&
            \mathbb{F}_{2}[V \setminus \{x\}]
                \arrow[d, dashed, "\psi","\sim"{swap,rotate=90, anchor=south}] \\
            \&
            \mathbb{F}_{2}[W]
        \end{tikzcd}
    \end{align}
\end{lemma}
\begin{proof}
    Since $\ker(\phi) = (x + Q)$ and
    $\Ima(\phi) = \mathbb{F}_{2}[V \setminus \{x\}]$,
    by the first isomorphism theorem for rings
    there is a unique isomorphism
    $\mathbb{F}_{2}[V]/(x + Q) \cong \mathbb{F}_{2}[V \setminus \{x\}]$,
    which gives the desired universal property.

    Now, suppose that
    $\phi':\mathbb{F}_{2}[V] \rightarrow \mathbb{F}_{2}[W]$
    is degree-nonincreasing and $\ker(\phi') = (x+Q)$.
    We notice that
    for $f \in \mathbb{F}_{2}[V]$, with $\deg(f) = 1$ for $c \in \mathbb{F}_{2}$,
    \begin{align}
        f + (x + Q) = c + (x + Q)
        \iff \phi(f) = c
        \iff \phi'(f) = c ~\label{equation_degree_one_constant_universal_iso}
    \end{align}
    so that $\deg(\phi(f)) = 0$ iff $\deg(\phi'(f)) = 0$.
    Since $\psi$ sends $\phi(f)$ to $\phi'(f)$,
    and all degree 1 polynomials of $\mathbb{F}_{2}[V \setminus \{x\}]$
    are in the image of degree 1 polynomials of $\mathbb{F}_{2}[V]$
    under $\phi$, it follows that $\psi$ is degree preserving.

    In addition, let $\phi'$ be simple.
    Then for all $v \in V \setminus \{x\}$, $\psi$ sends $\phi(v) = v$
    to $\phi'(v)$ which by assumption must have $\lvert \Var(\phi'(v)) \rvert = 1$,
    and so $\psi$ is simple.

    \EXTRA{
        Let $\alpha \in \ker(\phi)$
        be written as $\alpha(x) \in \mathbb{F}_{2}[V \setminus \{x\}][x]$.
        Then by pseudo-division by monic $(x + Q) \in \mathbb{F}_{2}[V \setminus \{x\}][x]$.
        We have
        \begin{align*}
            \alpha(x) = q(x)(x + Q) + r
        \end{align*}
        where $r \in \mathbb{F}_{2}[V \setminus \{x\}]$.
        It follows that
        \begin{align*}
            \phi(\alpha(x)) = r = 0
        \end{align*}
        and so $\alpha \in (x+Q)$. The converse is obvious so that $\ker(\phi) = (x +Q)$.
        It follows that $\mathbb{F}_{2}[V \setminus \{x\}] \cong \mathbb{F}_{2}[V]/(x+Q)$.

        Since every linear form in $\mathbb{F}_{2}[V \setminus \{x\}]$
        can be written as $f[x \leftarrow Q]$ for $f \in \mathbb{F}_{2}[V]$
        with $\deg(f) = 1$ (and likewise for $\mathbb{F}_{2}[V \setminus \{y\}]$),
        \begin{align*}
            \{ g \in \mathbb{F}_{2}[V \setminus \{x\}] \suchthat \deg(g) \le 1]\} 
            & = \{ f [x \leftarrow Q]  \in \mathbb{F}_{2}[V] \suchthat \deg(f) = 1]\} \\
            \{ g \in \mathbb{F}_{2}[V \setminus \{y\}] \suchthat \deg(g) \le 1]\} 
            & = \{ f [y \leftarrow Q']  \in \mathbb{F}_{2}[V] \suchthat \deg(f) = 1]\}
        \end{align*}
    }
\end{proof}

By Diagram~\ref{universal_rewrite_diagram}
which gives the universal property and conclusions on the induced
isomorphisms in Lemma~\ref{universal_rewrite},
we can make statements about the equivalence of path sums.
For example, consider $Q = y + Q_{1} = w + Q_{2}$ where $\deg(Q) = 1$,
$y \notin \Var(Q_{1})$, and $ w \notin \Var(Q_{2})$, and let
\begin{align*}
    A &= s_{A} \sum_{V_{A}}(-1)^{xQ + R} \ket{O}\bra{I}
\end{align*}
satisfy the precondition of the HH rule. Then by choice
of either $\phi = [y \leftarrow Q_{1}]$ or $\phi' = [w \leftarrow Q_{2}]$
we have two different HH rewrites $A \rightarrow B_{1}$ and $A \rightarrow B_{2}$
respectively. However, by Lemma~\ref{universal_rewrite}
we see that there exists an affine transformation $\psi$
such that $\psi(B_{1}) = B_{2}$. And furthermore,
if $\lvert \Var(Q) \rvert \leq 2$ then $\phi, \phi'$
are simple and thus so is $\psi$, and we have that $B_{1} \sim B_{2}$
under our definition of equivalence.
In particular this implies that the choice of target variable
in the HH rule yields unique path sums up to simple equivalence.
As a consequence, we will abuse notation to write $[Q \leftarrow 0]$
for a degree~1 polynomial $Q$ to mean $[y \leftarrow Q']$
for any variable $y \in \Var(Q)$, $Q = y + Q'$.
The following lemma stated in general for affine equivalence
intuitively says that if the affine transformation
between $A$ and $B$ is compatible with a ~\ref{eq:hh} rewrite
applied on $A$ in the sense that there is a corresponding
rewrite available on $B$, then the results are affine equivalent.

\begin{lemma}~\label{hh_simple}
    Let $A \in \PS$ have the form consistent with the precondition
    of the~\ref{eq:hh} rule, and let $A \rightarrow_{hh} A'$
    where $\deg(Q) = 1$.
    \begin{align*}
        A &= s_{A} \sum_{V_{A}}(-1)^{xQ + R} \ket{O}\bra{I} \\
        A' &= s_{A} \sum_{V_{A}\setminus \{x\} }
            (-1)^{R[Q \leftarrow 0]}
            \HOMSTATE{O}{I}{[Q \leftarrow 0]}
    \end{align*}
    Suppose $\phi$ is an affine transformation of $A$
    such that $B = \phi(A)$ admits an application of the~\ref{eq:hh} rule yielding
    \begin{align*}
        B' &= s_{A} \sum_{V_{B} \setminus \{z\}}
            (-1)^{\phi(R)[\phi(Q) \leftarrow 0]}
            \HOMSTATE{\phi(O)}{\phi(I)}{[\phi(Q) \leftarrow 0]}
    \end{align*}
    If $\phi (\mathbb{F}_{2}[V_{A} \setminus \{x\}]) \subseteq \mathbb{F}_{2}[V_{B} \setminus \{z\}]$,
    then there exists an affine transformation $\psi$
    such that $\psi(A') = B'$. In addition
    if $[Q \leftarrow 0]$ and $\phi$ are simple,
    then $\psi $ is simple.
\end{lemma}
\begin{proof}
    Letting $
        \phi': \mathbb{F}_{2}[V_{A} \setminus \{x\} ]
            \rightarrow \mathbb{F}_{2}[V_{B} \setminus \{z\} ]
    $ be the restriction of $\phi$,
    since $x \notin \Var(Q,R,O,I)$, it follows
    that
    \begin{align*}
        B' &= s_{A} \sum_{V_{A} \setminus \{z\}}
            (-1)^{\phi'(R)[\phi'(Q) \leftarrow 0]}
            \HOMSTATE{\phi'(O)}{\phi'(I)}{[\phi'(Q) \leftarrow 0]}
    \end{align*}
    By the universal property ~\ref{universal_rewrite} of $[Q \leftarrow 0]$
    we have that the affine transformation
    $\phi'$ induces an affine transformation
    $\psi$ such that the following diagram commutes.
    \[
        \begin{tikzcd}[column sep=2.5cm, row sep=huge]
            \mathbb{F}_{2}[V_{A'}]
                \arrow[r, "{[Q \leftarrow 0]}"]
                \arrow[d, "\phi'"']
            & 
            \mathbb{F}_{2}[V_{A'}]/(Q)
                \arrow[d, "\psi", dashed] \\
            \mathbb{F}_{2}[V_{B'}]
                \arrow[r, "{[\phi'(Q) \leftarrow 0]}"']
            &
            \mathbb{F}_{2}[V_{B'}]/(\phi'(Q))
        \end{tikzcd}
    \]
    Where $\mathbb{F}_{2}[V_{A'}]/(Q)$ and
    $\mathbb{F}_{2}[V_{B'}]/(\phi'(Q))$ are understood
    as the codomain of the substitution homomorphism for arbitrary choice of
    variable as in Lemma~\ref{universal_rewrite}.
    Furthermore if $\phi $, and $[Q \leftarrow 0]$ are simple,
    then $[\phi'(Q) \leftarrow 0]$ is simple
    and by Lemma~\ref{universal_rewrite}, it follows
    that $\psi$ is simple.
\end{proof}

The $\beta $ property follows from the fact that a simple equivalence
$A \sim B$,
is always compatible with~\ref{eq:hh} rewrites. This is to say
that there is always a corresponding rewrite on $B$ to any given ~\ref{eq:hh}
rewrite on $A$, and Lemma~\ref{hh_simple} guarantees that the results
are simple equivalent.

\begin{corollary}~\label{beta_hh}
    The ~\ref{eq:hh} rule satisfies property $\beta $
    of Definition~\ref{local_confluence_modulo_equivalence}.
\end{corollary}
\begin{proof}
    Let $B = \phi(A)$ where $\phi$ is a simple transformation.
    Then $\phi(x) = z + c$, for some $ z \in V_{B}, c \in \mathbb{F}_{2}$, and
    \begin{align*}
        B &= s_{A} \sum_{V_{B}}
            (-1)^{z \phi(Q) + c \phi(Q) + \phi(R)} \ket{\phi(O)}\bra{\phi(I)}
    \end{align*}
    admits an HH rewrite of the form required by Lemma~\ref{hh_simple}.
    Specifically, the rewrite with pivot variable $z$
    yielding the homomorphism $[\phi(Q) \leftarrow 0]$.
\end{proof}

\begin{example}[]
    Consider the equivalent path sums from Example
    \ref{example_hh_simple_equivalence}
    where the LHS admits the first rewrite from Example
    \ref{example_hh_rule_simplifying_two_hadamards}.
    Corollary~\ref{beta_hh} ensures that there is a corresponding
    rewrite to the RHS and furthermore that their results are equivalent.
    \begin{center}
        \begin{tikzpicture}[>=Stealth, thick]
            \node (Y) at (-4.25, 0.5) {$
                \frac{1}{4}\sum_{\{u,w,x,y,z\} }
                    (-1)^{zw + xy + \textcolor{color1}{u}(w + y)}
                    \ket{z} \bra{x}
            $};
            \node (Z) at (4.25, 0.5) {$
                \frac{1}{4}\sum_{\{u,w,x,y,z\} }
                    (-1)^{xw + zy + z + \textcolor{color1}{u}(w + y + 1)}
                    \ket{x+1} \bra{z+1}
            $};
            \node (Yp) at (-4.25, -2) {$
                \frac{1}{4}\sum_{\{w,x,y,z\} }
                    (-1)^{zy + xy }
                    \ket{z} \bra{x}
            $};
            \node (Zp) at (4.25, -2) {$
                \frac{1}{4}\sum_{\{w,x,y,z\} }
                    (-1)^{xw + zw}
                    \ket{x+1} \bra{z+1}
            $};

            \draw[->] (Y) -- (Yp) node[midway, left] { {\footnotesize $
                \quad [w \leftarrow y]
            $ }
            };
            \draw[->, dashed] (Z) -- (Zp) node[midway, right] { {\footnotesize $
                [y \leftarrow w + 1] \quad
            $ }
            };

            \draw[-,
                decorate,
                decoration={snake, amplitude=0.5mm, segment length=8mm}]
                (Yp) to (Zp);
            \draw[draw=none] (Yp) -- (Zp) node[midway, above] {{\footnotesize $
                \psi
            $} };
            \draw[-,
                decorate,
                decoration={snake, amplitude=0.5mm, segment length=8mm}]
                (Y) to (Z);
            \draw[draw=none] (Y) -- (Z) node[midway, above] {{\footnotesize $
                \phi
            $} };
        \end{tikzpicture}
    \end{center}
    The resulting path sums can be seen to be equivalent by the
    simple transformation $\psi$ defined by
    \begin{align*}
        \psi:
        w \mapsto y, \quad
        x \mapsto z+1, \quad
        y \mapsto w, \quad
        z \mapsto x+1
    \end{align*}
    as a direct consequence of the unique isomorphism
    $\bar{\psi}:\mathbb{F}_{2}[x,y,z]
        \rightarrow \mathbb{F}_{2}[w,x,z]$
    such that
    \begin{align*}
        &\bar{\psi} \circ [w \leftarrow y]
        = [y \leftarrow w+1]\circ \phi \\
        &\text{defined by } \quad
        \bar{\psi}:
        x \mapsto z+1, \quad
        y \mapsto w, \quad
        z \mapsto x+1
    \end{align*}
\end{example}

Finally to prove the $\alpha$ property, we will further borrow
from ring theory to consider the case of a sequence
of $ \rightarrow_{hh}$. To this end, recall the following
which is a consequence of the third isomorphism theorem for rings.
For a ring $R$ with ideals $I,J \subseteq R$,
let $\pi: R \rightarrow R/I$ be the quotient homomorphism.
Then $\pi(J) = \pi(I+J) = (I+J)/I$ is an ideal of $R/I$ and
\begin{align}
    \frac{R/I}{\pi(J)} =
    \frac{R/I}{(I+J)/I} \cong \frac{R}{I + J}~\label{ring_hom_chain}
\end{align}
which says that if $\rho: R/I \rightarrow \frac{R/I}{\pi(J)}$
is again the quotient homomorphism, then
$\rho \circ \pi$ has the universal property of the quotient with respect
to ideal $I + J = \{i + j \suchthat i \in I, j \in J\}$.

Since we've seen that the substitution homomorphism
$[Q \leftarrow 0]$ can be identified with $\pi$ for $I = (Q)$ up to an affine equivalence,
we see that this allows for the universal property to extend to
a sequence of multiple rewrites. To see how this applies to
substitution homomorphisms, consider the following
which in particular implies that
for degree 1 polynomials $Q_{1}, Q_{2}$,
we have $Q_{1} [Q_{2} \leftarrow 0] = c$ iff $Q_{2} [Q_{1} \leftarrow 0] = c$.

\begin{lemma}~\label{substitution_mirror}
    Let $Q_{1}, Q_{2} \in \mathbb{F}_{2}[V]$ be polynomials
    such that $\deg(Q_{1}) = \deg(Q_{2}) = 1$.
    Then for $c \in \mathbb{F}_{2}$,
    $Q_{1}[Q_{2} \leftarrow 0] = c$ iff
    $Q_{1} + Q_{2} = c$.
\end{lemma}
\begin{proof}
    In Appendix~\ref{appendix_multilinear_forms}
\end{proof}

Now let $Q_{1}, Q_{2} \in \mathbb{F}_{2}[V]$ such that $\deg(Q_{1} + Q_{2}) = 1$,
so that $Q_{2}' = Q_{2}[Q_{1} \leftarrow 0]$ has degree $1$ by the above.
Then letting $V'' \subsetneq V' \subsetneq V$ be the variable sets with
the target variables of $[Q_{2}' \leftarrow 0],[Q_{1} \leftarrow 0]$
removed respectively, we have a composition of homomorphisms
\[
        \begin{tikzcd}[column sep=2cm, row sep=large, ampersand replacement=\&]
            \mathbb{F}_{2}[V]
                \arrow[r, "{[Q_{1} \leftarrow 0]}"]
                \&
            \mathbb{F}_{2}[V']
                \arrow[r, "{[Q_{2}' \leftarrow 0]}"]
                \&
            \mathbb{F}_{2}[V'']
        \end{tikzcd}
\]
By Lemma~\ref{universal_rewrite}, and Equation~\ref{ring_hom_chain} we have that
\begin{align*}
    \mathbb{F}_{2}[V'']
    \cong \frac{\mathbb{F}_{2}[V']}{(Q_{2}')}
    \cong \frac{\mathbb{F}_{2}[V]/(Q_{1})}{(\pi(Q_{2}))}
    \cong \frac{\mathbb{F}_{2}[V]}{(Q_{1}, Q_{2})}
\end{align*}
where $\pi: \mathbb{F}_{2}[V] \rightarrow \mathbb{F}_{2}[V]/(Q_{1})$
is the quotient map. Thus the homomorphism
\begin{align}
    \phi= [Q_{1} \leftarrow 0][Q_{2}' \leftarrow 0]
    ~\label{equation_multiple_rewrites}
\end{align}
satisfies the universal property
of the quotient
$\mathbb{F}_{2}[V] \rightarrow  \frac{\mathbb{F}_{2}[V]}{(Q_{1}, Q_{2})}$.
Furthermore, the same consequences about the induced isomorphism $\psi$ of Diagram
~\ref{figure_chain_universal_rewrite_diagram} follow, by the same proof as
Lemma~\ref{universal_rewrite}. Indeed, consider an alternative
$\phi': \mathbb{F}_{2}[V] \rightarrow \mathbb{F}_{2}[W]$
with $\ker(\phi') = \left( Q_{1}, Q_{2} \right)$,
then there is a unique isomorphism
$\psi: \mathbb{F}_{2}[V''] \rightarrow \mathbb{F}_{2}[W]$
such that the following commutes
\begin{align}~\label{figure_chain_universal_rewrite_diagram}
    \begin{tikzcd}[wire types={n,n}, column sep=2.5cm,
        row sep=large, ampersand replacement=\&]
        \mathbb{F}_{2}[V]
            \arrow[r, "{\phi}"]
            \arrow[dr, "\phi'"] \&
        \mathbb{F}_{2}[V'']
            \arrow[d, dashed, "\psi","\sim"{swap,rotate=90, anchor=south}] \\
        \&
        \mathbb{F}_{2}[W]
    \end{tikzcd}
\end{align}
and furthermore, we have by the exact same reasoning as in
Equation~\ref{equation_degree_one_constant_universal_iso} of
Lemma~\ref{universal_rewrite},
that if $\phi'$ is degree-preserving (respectively, simple) then
$\psi$ is degree-preserving (respectively, simple).

This observation about chaining homomorphisms
can be extended further, and in particular guarantees affine or simple equivalence
of path sums which are results of different sequences of $ \rightarrow_{hh}$ so
long as they have the same kernel. We use this reasoning to prove the $\alpha$
property by a case analysis of each pair of possible rewrites. The unification
of path sums will involve sequences of rewrites, which in the view developed,
will correspond to a composition of homomorphisms which is yet another
homomorphism. Then by observing when two seemingly distinct rewrite
sequences have the same kernel, we can reason by
the universal property of the quotient that the results
are in fact simple equivalent.

\begin{lemma}~\label{proposition_alpha_property}
    The rewrite system $ \rightarrow $ of Definition~\ref{rewrite_system}
    on $\PS$ satisfies the $\alpha$ property.
\end{lemma}
\begin{proof}
    Let $A \rightarrow B$ and $A \rightarrow C$ as in top two arrows in the alpha property in
    Figure~\ref{local_confluence_modulo_equivalence_figure}.
    We consider cases of each rewrite in
    $\{\rightarrow_{elim}, \rightarrow_{hh}, \rightarrow_{z}\} $
    and by symmetry need only consider each unordered pair.
    Whenever one of the rewrites is $ \rightarrow_{elim}$,
    we can trivially obtain a common $D \in \PS$, since variable elimination
    and all other rewrites commute.
    So we consider only the pairs of rewrites taken from
    $\{\rightarrow_{hh}, \rightarrow_{z}\} $.
    The case when both $A \rightarrow_{z} B$, $A \rightarrow_{z} C$
    is trivial as $B = C = \Null$.
    So we show the remaining two cases.

    \pfcase{1}{$A \rightarrow_{z} B$ and $A \rightarrow_{hh} C$}
    Let $x$ be the pivot variable for $A \rightarrow_{hh} C$,
    and $z$ be the variable of $A$ as in $z$ in the LHS for $A \rightarrow_{z} B = \Null$.
    Then $x$ and $z$ must be distinct variables and $z \notin \Var(Q)$
    where $A, C$ have general forms
    \begin{align*}
        A &= \sum_{V_{A}} (-1)^{(
            \var{z}
            + \var{x}\bool{Q} + \rool{R}
        )} \ket{O_{A}}\bra{I_{A}}\\
        C &= \sum_{V_{A} \setminus \{x\} } (-1)^{(
            \var{z}+ \rool{R}[Q \leftarrow 0]
        )}
        \HOMSTATE{O_{A}}{I_{A}}{[\bool{Q} \leftarrow 0]}
    \end{align*}
    Where $x, z \notin \Var(Q, R, O_{A}, I_{A})$.
    Then $C$ rewrites to $B= \Null$ by an application of $\rewrite_{z}$.

    \pfcase{2}{$A \rightarrow_{hh} B$ and $A \rightarrow_{hh} C$}
    We distinguish cases on the ``pivot'' variable $x$ as
    in the HH rule of Definition ~\ref{rewrite_system}.
    We let $x,w$ be the ``pivot'' variables for
    $A \rightarrow B$ and $A \rightarrow C$ respectively.
    This is to say that $A$ can be expressed in two forms,
    \begin{align*}
        A
        = s_{A} \sum_{V_{A}} (-1)^{xQ_{1} + R_{1}} \ket{O_A} \bra{I_A}
        = s_{A} \sum_{V_{A}} (-1)^{wQ_{2} + R_{2}} \ket{O_A} \bra{I_A}
    \end{align*}
    both satisfying the precondition of the HH rule.

    If $x = w$, then we have that $Q_{1} = Q_{2}$ so that
    both $A \rightarrow B$, and $A \rightarrow C$ induce
    substitution homomorphism
    $[Q_{1} \leftarrow 0] $, which by Lemma~\ref{universal_rewrite}, implies that $B \sim C$.

    Now suppose that $x \neq w$.
    We notice that $x \in Var(Q_{2}) \iff w \in Var(Q_{1})$ since both sides
    imply that $xw$ is a term of $P_{A}$.
    Consider the case where $x \notin Var(Q_{2})$.
    It follows that $A$ must have the general form
    \begin{align*}
        A = s_{A} \sum_{V_{A}} (-1)^{xQ_{1} + wQ_{2} + R} \ket{O_A} \bra{I_A}
    \end{align*}
    where $x, w \notin Var(Q_{1}, Q_{2}, R, O_{A}, I_{A})$.
    It follows that B, C have forms
    \begin{align*}
        B &= s_{A} \sum_{V_{A} \setminus \{x\} }
            (-1)^{wQ_{2}' + R[Q_{1} \leftarrow 0]}
            \HOMSTATE{O_{A}}{I_{A}}{[Q_{1} \leftarrow 0]} \\
        C &= s_{A} \sum_{V_{A} \setminus \{w\} }
            (-1)^{xQ_{1}' + R[Q_{2} \leftarrow 0]}
            \HOMSTATE{O_{A}}{I_{A}}{[Q_{2} \leftarrow 0]}
    \end{align*}
    where $Q_{2}' = Q_{2}[Q_{1} \leftarrow 0]$
    and $Q_{1}' = Q_{1}[Q_{2} \leftarrow 0]$.
    By Lemma ~\ref{substitution_mirror}, it follows that for $c \in \mathbb{F}_{2}$,
    \begin{align*}
        Q_{2}' = c \iff Q_{1}' = c \iff Q_{2} + Q_{1} = c
    \end{align*}
    Thus if $Q_{2}' = 0$, then $Q_{1}' = 0$
    and $Q_{1} = Q_{2}$, so $B \sim C$ by simply mapping $w \mapsto x$.
    If $Q_{2}' = 1$, then $Q_{1}' = 1$
    and both $B, C$ rewrite to $\Null$ by $ \rightarrow_{z}$.
    So we are left with the case that $\deg(Q_{2}') = \deg(Q_{1}') = 1$.
    Since $\lvert \Var(Q_{1}) \rvert \leq 2, \; \lvert \Var(Q_{2}) \rvert \leq 2$,
    and $[Q_{2} \leftarrow 0], \; [Q_{1} \leftarrow 0]$ are simple,
    it follows that $\lvert \Var(Q_{1}') \rvert \leq 2, \; \lvert \Var(Q_{2}') \rvert \leq 2$.
    Thus $B$ and $C$ rewrite to $B'$ and $C'$ respectively
    \begin{align}
        B' &= \left(  s_{A} \sum_{V_{A} \setminus \{x,w\} }
            (-1)^{R}
            \ket{O_A} \bra{I_A} \right)
            [Q_{1} \leftarrow 0][Q_{2}' \leftarrow 0]
            ~\label{equation_alpha_hh_1} \\
        C' &= \left(  s_{A} \sum_{V_{A} \setminus \{x,w\} }
            (-1)^{R}
            \ket{O_A} \bra{I_A} \right)
            [Q_{2} \leftarrow 0][Q_{1}' \leftarrow 0]
            ~\label{equation_alpha_hh_2}
    \end{align}
    Since both $\phi_{1} = [Q_{1} \leftarrow 0][Q_{2}' \leftarrow 0],
        \phi_{2} = [Q_{2} \leftarrow 0][Q_{1}' \leftarrow 0]
    $ have the same kernel $(Q_{1}, Q_{2})$, it follows
    by the discussion of ~\ref{equation_multiple_rewrites},
    that $B', C'$ are affinely related
    by some $\phi(B') = C'$, and furthermore $\phi$ is simple since both
    $\phi_{1},\phi _{2}$ are simple.

    To finish the proof, we consider lastly when $x \in \Var(Q_{2})$.
    It follows that $A$ has the general form
    \begin{align*}
        A = s_{A} \sum_{V_{A}} (-1)^{xQ_{1}' + wQ_{2}' + xw + R} \ket{O_A} \bra{I_A}
    \end{align*}
    where $x, w \notin Var(Q_{1}', Q_{2}', R, O_{A}, I_{A})$.
    Then by Lemma ~\ref{universal_rewrite}, up to a simple equivalence,
    we can let $A \rightarrow B$, $A \rightarrow C$
    induce $[w \leftarrow Q_{1}'], [x \leftarrow Q_{2}'] $ respectively.
    Thus we have
    \begin{align*}
        B \sim s_{A} \sum_{V_{A} \setminus \{x\} }
            (-1)^{Q_{1}'Q_{2}' + R}
            \ket{O_A} \bra{I_A}
        \quad
        \quad
        \quad
        C \sim s_{A} \sum_{V_{A} \setminus \{w\} }
            (-1)^{Q_{1}'Q_{2}' + R}
            \ket{O_A} \bra{I_A}
    \end{align*}
    where the RHS of each equivalence are equivalent by the isomorphism which sends
    $w \mapsto x$.
\end{proof}

The following example illustrates the case of the $\alpha$
property expressed by~\Cref{equation_alpha_hh_1,equation_alpha_hh_2},
where the universal property of the homomorphism of the form
~\ref{equation_multiple_rewrites}, ensures that the resulting
path sums are simple equivalent.
\begin{example}[]
    Consider the two potential rewrites due to the~\ref{eq:hh} rule
    with respect to pivot variables $x_{0}$ and $y_{1}$ respectively.
    They can be seen to be unified by \ref{eq:hh} rule applications
    with pivots $y_{1}$ and $x_{0}$ respectively.
    \begin{center}
        \begin{tikzpicture}[>=Stealth, thick]
            \node (X) at (0, 3) { $
                \frac{1}{8}
                \sum_{\bigcup_{i \in [2] }\{x_{i},y_{i},z_{i}\}}
                (-1)^{
                    \textcolor{color1}{x_{0}}(x_{1}+y_{0}+1)
                    + \textcolor{color2}{y_{1}}(z_{1} + x_{1} +1)
                    + y_{0}z_{0} }
                \ket{z_{0}, z_{1}}
            $};
            \node (Y) at (-4.5, 0.5) {$
                \frac{1}{8}
                \sum_{\{x_{1}\} \cup \bigcup_{i \in [2] }\{y_{i},z_{i}\}}
                (-1)^{
                    \textcolor{color2}{y_{1}}(z_{1} + y_{0})
                    + y_{0}z_{0} }
                \ket{z_{0}, z_{1}}
            $};
            \node (Z) at (4.5, 0.5) {$
                \frac{1}{8}
                \sum_{\{y_{0}\} \cup \bigcup_{i \in [2] }\{x_{i},z_{i}\}}
                (-1)^{
                    \textcolor{color1}{x_{0}}(x_{1}+y_{0}+1) + y_{0}z_{0} }
                \ket{z_{0}, x_{1} + 1}
            $};
            \node (Yp) at (-3.5, -2) {$
                \frac{1}{8}
                \sum_{\{x_{1},y_{0},z_{0},z_{1}\} }
                (-1)^{y_{0}z_{0} }
                \ket{z_{0}, y_{0}}
            $};
            \node (Zp) at (3.5, -2) {$
               \frac{1}{8}
                \sum_{\{x_{1},y_{0},z_{0},z_{1}\} }
                (-1)^{x_{1}z_{0} + z_{0} }
                \ket{z_{0}, x_{1} + 1}
            $};

            \draw[->] (X) -- (Y) node[midway, left] { {\footnotesize $
                [x_{1} \leftarrow y_{0} + 1] \quad
            $ } };
            \draw[->] (X) -- (Z) node[midway, right] { {\footnotesize $
                \quad [z_{1} \leftarrow x_{1} + 1]
            $ } };

            \draw[->, dashed] (Y) -- (Yp) node[midway, left] { {\footnotesize $
                \quad [z_{1} \leftarrow y_{0}]
            $ }
            };
            \draw[->, dashed] (Z) -- (Zp) node[midway, right] { {\footnotesize $
                [y_{0} \leftarrow x_{1} + 1] \quad
            $ }
            };

            \draw[-,
                decorate,
                decoration={snake, amplitude=0.5mm, segment length=8mm}]
                (Yp) to (Zp);
            \draw[draw=none] (Yp) -- (Zp) node[midway, above] {{\footnotesize $
                \phi
            $} };
        \end{tikzpicture}
    \end{center}
    By the fact that the composite homomorphisms down each path
    have the same kernel $(x_{1} + y_{0} +1, z_{1} + x_{1} +1)$,
    they are related by the simple transformation $\phi$
    that has the following action,
    \begin{align*}
        \phi:
        x_{1} \mapsto y_{0},
        \quad
        y_{0} \mapsto x_{1} + 1,
        \quad
        z_{0} \mapsto z_{0},
        \quad
        z_{1} \mapsto z_{1}
    \end{align*}
    which is a direct consequence of the unique isomorphism
    $\bar{\phi}:
        \mathbb{F}_{2}[y_{0}, z_{0}]
        \rightarrow \mathbb{F}_{2}[x_{1}, z_{0}]$
    such that
    \begin{align*}
        &\bar{\phi} \circ ([z_{1} \leftarrow y_{0}][x_{1} \leftarrow y_{0} + 1])
        = [y_{0} \leftarrow x_{1} + 1][z_{1} \leftarrow x_{1} + 1] \\
        &\text{defined by } \quad
        \bar{\phi}: y_{0} \mapsto x_{1} + 1,
        \quad
        z_{0} \mapsto z_{0}
    \end{align*}
\end{example}

Now, combining Lemma~\ref{proposition_alpha_property}
which proves the $\alpha$ property,
and Lemmas~\ref{beta_z},~\ref{beta_elim},~\ref{beta_hh}
which prove the $\beta$ property,
we have the following theorem.

\begin{theorem}\label{thm:confluence}
    The rewrite system $\rewrite$ of Definition~\ref{rewrite_system}
    on $\PS$ is confluent modulo simple equivalence~$\sim$.
\end{theorem}

\moduleclose

\section{Polynomial-Time Simulation of Hidden Shift Circuits}\label{sec:simulation}
Having shown that a circuit $C$ over gate set $\mathcal{G} = \GATESET$
can be interpreted
as a path sum $\llbracket C \rrbracket$ in polynomial time in the volume of $C$,
and that the rewrite system $ \rightarrow $ to simplify $\llbracket C \rrbracket$
is confluent, it remains to determine how confluence of the
rewrite system ~\ref{rewrite_system} affects the time complexity of the
simplification
and evaluation phases of algorithms~\ref{alg:strong_sim}
and~\ref{alg:bqp_sim}.
From Corollary~\ref{polynomial_rewrites}, we have on input circuit $C$,
that the number of rewrites is bounded by $\poly(\lvert C \rvert)$
where $\lvert C \rvert $ is the volume of circuit $C$.
Furthermore,
each rewrite takes at most
polynomial time in the size of path sum at the beginning of the iteration
to identify and apply by a linear scan of the path sum expression.
By the discussion at the end of~\cref{section_simulation},
we have that the size of the path sum remains $\poly(\lvert C \rvert)$
throughout repeated rewrites, so that the path sum at the beginning
of each iteration is in fact of size $\poly(\lvert C \rvert)$.
By combining these facts together, namely that we have $\poly(\lvert C \rvert)$
many iterations, each taking $\poly(\lvert C \rvert)$ time,
we have that the while loop in algorithms \ref{alg:strong_sim}
and~\ref{alg:bqp_sim} terminate in $\poly(\lvert C \rvert)$ time.
Thus, for the algorithms
to be efficient, it is sufficient for the final
evaluation to take polynomial time. This is achieved when the path sum
$\llbracket C \rrbracket $ can be simplified to a path sum of sufficiently
few variables. As the preceding discussion suggests,
the notion of efficiency for a family of circuits will be polynomial-time
in the volume of each circuit.

\begin{lemma}~\label{strong_sim_lemma}
    Let $\mathcal{C}$ be a family of circuits over $\mathcal{G}$. For circuit $C \in \mathcal{C}$
    with signature $n \rightarrow l$,
    $\vec{x} \in \mathbb{F}_{2}^{n}$,
    and $\vec{y} \in \mathbb{F}_{2}^{l}$,
    if
    $
        \bra{\vec{y}} \circ  \llbracket C \rrbracket \circ \ket{\vec{x}}
        \overset{\ast }{ \rightarrow } A
    $
    such that $\lvert V_{A} \rvert = O(\log \lvert C \rvert )$,
    then algorithm~\ref{alg:strong_sim} outputs $\bra{\vec{y}} C \ket{\vec{x}}$
    in polynomial time.
\end{lemma}
\begin{proof}
    Construction of $\llbracket C \rrbracket$ takes time
    and has size $O(\lvert C \rvert^{m+2})$.
    For at most $O(\lvert C \rvert^{m+2})$ times,
    rewrites are applied which each take $O(\lvert C \rvert^{m+2})$ time.
    After rewriting to a normal form $A'$ which is $A' \sim A$,
    we have evaluation on $2^{O(\log \lvert C \rvert) }$ points.
    Let $\lvert V_{A'} \rvert   = \lvert V_{A} \rvert \leq d \log \lvert C \rvert $
    asymptotically.
    Then we have at most $\lvert C \rvert^{d}$ points, each polynomial evaluation
    takes $O(\lvert V_{A} \rvert^{m+1})$ time.
    So the total time for evaluating $A'$~is
    \begin{align*}
        &\lvert C \rvert^{d} O(\lvert V_{A} \rvert^{m + 1})
        = O(\lvert C \rvert^{m + d + 1})
    \end{align*}
    The time complexity for algorithm is then
    \begin{align*}
        O(\lvert C \rvert^{2m + 4})
        + O(\lvert C \rvert^{m + d + 1})
        = O(\lvert C \rvert^{2m + d + 4})
    \end{align*}
\end{proof}

To state a similar result for algorithm~\ref{alg:bqp_sim},
it will be useful to state a lemma
relating rewrites and path sum compositions and tensor products.
In the following, let $\overset{\epsilon}{ \rightarrow }$
be the reflexive closure of the binary relation $ \rightarrow $.

\begin{lemma}~\label{compositional}
    Let $A, B\in \PS$, be path sums such
    that $A \overset{\epsilon}{\rightarrow} A'$
    and $B \overset{\epsilon}{\rightarrow} B'$.
    Then whenever the following are well-defined,
    \begin{align}
        A \circ B
        &\overset{\ast }{\rightarrow}
        A' \circ B' ~\label{equation_compositional_1} \\
        A \otimes B
        &\overset{\ast }{\rightarrow}
        A' \otimes B' ~\label{equation_compositional_2} \\
        A^{\dagger}
        &\overset{\epsilon }{\rightarrow}
        \left( A' \right) ^{\dagger} ~\label{equation_compositional_3}
    \end{align}
    where the RHS of ~\Cref{equation_compositional_1,equation_compositional_2}
    is $0$ in the exceptional case where either $A \rightarrow_{z} 0$
    or $B \rightarrow_{z} 0$.
\end{lemma}
\begin{proof}
    In Appendix~\ref{appendix_composition}
\end{proof}

\begin{lemma}~\label{bqp_sim_lemma}
    Let $\mathcal{C}$ be a family of circuits over $\mathcal{G}$.
    For circuit $C \in \mathcal{C}$
    on $n$ qubits, and $\vec{x} \in \mathbb{F}_{2}^{n}$,
    if~$
        \llbracket C \rrbracket \circ \ket{\vec{x}}
        \overset{\ast }{ \rightarrow } A
    $
    such that $\lvert V_{A} \rvert = O(\log \lvert C \rvert )$,
    then for all $i \in [n]$, algorithm~\ref{alg:bqp_sim} terminates in
    polynomial time on input $(C, \vec{x}, i)$.
\end{lemma}
\begin{proof}
    Without loss of generality let $i = 1$. The path sum $g$ in
    algorithm~\ref{alg:bqp_sim} is constructed in time and has size $O(\lvert C \rvert^{m+2})$.
    Then path sum $f$ of the algorithm has size at most
    \begin{align*}
        2O(\lvert C \rvert^{m+2}) + O(n)  = O(\lvert C \rvert^{m+2})
    \end{align*}
    thus simplification of $f$ consists of at most
    $O(\lvert C \rvert^{m+2})$ rewrites each taking $O(\lvert C \rvert^{m+2})$
    time, which takes time $ O(\lvert C \rvert^{2m+4})$.
    By assumption $g \overset{\ast }{\rightarrow} A$
    such that $\lvert V_{A} \rvert = O(\log \lvert C \rvert)$
    so that $\lvert V_{A} \rvert \leq d \log \lvert C \rvert$ asymptotically
    for some constant $d$.
    It follows by Lemma~\ref{compositional} that
    $g^{\dagger } \overset{\ast}{\rightarrow} A^{\dagger}$
    which has the same number of variables as $A$.
    Let $B = A^{\dagger}$ to simplify notation.
    By Lemma~\ref{compositional}, it follows that for path sum $f$ of the algorithm
    \begin{align*}
        f &\overset{\ast }{ \rightarrow } B
        \circ \left(\ket{1}\bra{1} \otimes I_{n-1} \right)
        \circ A \\
        & = B \circ \frac{s_{A}}{2^n}\sum_{\substack{
            V_{A} \cup \{x_{2}, \dots, x_{n}\} \\
            \cup \{z, y_{2}, \dots, y_{n}\}
            }} (-1)^{P_{A} + z(1 + O_{A, 1}) + \sum y_{i}(x_{i} + O_{A,i})}
            \ket{1, x_{2}, \dots, x_{n}} \\
        & \overset{\ast }{ \rightarrow } B \circ \frac{s_{A}}{2}\sum_{\substack{
            V_{A} \cup \{z\}
            }} (-1)^{P_{A} + z(1 + O_{A, 1})}
            \ket{1, O_{A,2}, \dots, O_{A,n}} \\
        & =\frac{s_{B} s_{A}}{2^{n+1}}\sum_{\substack{
            V_{A} \cup V_{B} \cup \{z\} \\
            \cup \{w, y_{2}, \dots, y_{n}\}
        }} (-1)^{P_{A} + z(1 + O_{A, 1}) + P_{B} + w(1 + I_{B,1}) s+ \sum y_{i}(I_{B,i} + O_{A,i})}
            \ket{}\bra{} 
    \end{align*}
    where $\lvert \Var(I_{B,i} + O_{A,i}) \rvert \leq 2$ for $i = 2,\dots,n$ so that either
    an HH, Elim, or Z rule can be applied to remove each $y_{i}$.
    Thus we have that $f \overset{\ast}{\rightarrow} C$ such that
    $V_{C} = V_{A} \cup V_{B} \cup  \{w, z\}$, so that
    $\lvert V_{C} \rvert \leq 2d \log \lvert C \rvert + 2 = O(\log \lvert C \rvert )$.
    Let $\lvert V_{C} \rvert \leq d' \log \lvert C \rvert $ asymptotically.
    Then after the simplification loop of the algorithm,
    $f \sim C$ and so it has the same number of variables.
    Thus evaluation occurs on at most $2^{d' \log \lvert C \rvert} = \lvert C \rvert^{d'}$
    points for a polynomial with at most $O(\lvert V_{C} \rvert^{m+1})$ terms.
    Each term takes $m$ multiplications which is a constant, so that
    evaluation takes
    \begin{align*}
        \lvert C \rvert^{d'} O(\lvert V_{C} \rvert^{m+1})
        = O(\lvert C \rvert^{d'} \lvert C \rvert^{m+1})
        = O(\lvert C \rvert^{m + d' + 1})
    \end{align*}
    Thus, algorithm ~\ref{alg:bqp_sim} terminates in time
    \begin{align*}
        O(\lvert C \rvert^{2m + 4}) +
        O(\lvert C \rvert^{m + d' + 1})
        = O(\lvert C \rvert^{2m + d' + 4})
    \end{align*}
\end{proof}

Finally we prove our main result, that the family of hidden shift
circuits $C_{(\pi, g, \vec{s})}$ for Roetteler's shifted bent function algorithm
is polynomial-time simulable via rewrites of its path sum.
For a hidden shift circuit $C_{(\pi, g, \vec{s})}$, its classical
simulation amounts to applying algorithm ~\ref{alg:bqp_sim} on
\begin{align*}
    \left( C_{(\pi, g, \vec{s})}, \vec{0}, i\right)
\end{align*}
for each $i \in [n]$ to build the shift $\vec{s}$.
To prove that Algorithm ~\ref{alg:bqp_sim} simulates
hidden shift circuits efficiently, by Lemma~\ref{bqp_sim_lemma},
it suffices to prove that
$\llbracket C_{(\pi, g, \vec{s})} \rrbracket  \circ \ket{00 \dots 0}$
simplifies to sufficiently few variables. In the following proposition,
we will see that in fact, all variables can be removed by virtue of the
proof of correctness seen in~\cref{sec:prelim}. This is to say
that the proof of correctness can essentially be reproduced formally
with the rewrite system $\rightarrow$. Then by confluence of the rewrite system
$ \rightarrow $, ~\cref{alg:bqp_sim}, is guaranteed to sufficiently simplify the corresponding
path sum by applying rewrites in a structure-oblivious manner,
to yield the $i$-th bit of the shift in polynomial time.

\begin{proposition}~\label{hiddenshift_rewrite_existence}
    Given a hidden shift circuit $C_{(\pi, g, \vec{s})}$,
    the path sum $\llbracket C_{(\pi, g, \vec{s})} \rrbracket \circ \ket{00 \dots 0} $
    can be computed and has size polynomial in the volume
    $\lvert  C_{(\pi, g, \vec{s})}\rvert $ of
    $C_{(\pi, g, \vec{s})}$. Furthermore
    with the rewrite system $\rewrite $ of Definition~\ref{rewrite_system},
    \begin{align*}
        \llbracket C_{(\pi, g, \vec{s})} \rrbracket \circ \ket{00 \dots 0}
        \stackrel{\ast}{\rewrite} \ket{\vec{s}}
    \end{align*}
\end{proposition}
\begin{proof}
    We first note that through at most $\lvert  C_{(\pi, g, \vec{s})}\rvert $
    applications of $ \rightarrow_{hh}$,
    the path sum
   $\llbracket C \rrbracket \circ \ket{00 \dots 0}$ may be re-written to the form
   \[
    \frac{1}{2^{3n}}\sum_{x,y,z}(-1)^{g(x) + \langle x,y\rangle + \overline{f}(y) + \langle y, z\rangle}\ket{z}
   \]
    as in the proof of correctness in \cref{sec:prelim}, where the phase polynomial is explicitly expanded to algebraic normal form. Now it can be observed that the proof of correctness given in \cref{sec:prelim} follows from $3n$ applications of $ \rightarrow_{hh}$: the first sequence of $n$ rewrites pivots on variables of $x_1$ giving the simple substitutions $[y_1 \gets \phi(x_2) + \pi(s_2)] $, while the remaining $2n$ independent rewrites correspond to pivots $y_2$ and $x_2$ with substitutions $[z_2 \gets s_2] $ and $[z_1 \gets s_1] $
Since the proof ends in $\ket{s_1,s_2} = \ket{s}$, we have the desired result.
\end{proof}

Proposition~\ref{hiddenshift_rewrite_existence} ensures the existence of a rewrite
sequence resulting in a path sum of sufficiently few variables. Thus by applying
Lemma ~\ref{bqp_sim_lemma}, we can now derive the main result of our paper,
namely that hidden shift
circuits are simulated by Algorithm ~\ref{alg:bqp_sim} in polynomial time.

\begin{corollary}
    Let $\mathcal{C} = \{C_{(\pi, g, \vec{s})}\} $ be the hidden shift circuit family
    of Definition~\ref{bravyi-gosset-roetteler circuits}.
    Then for each $C_{(\pi, g, \vec{s})} \in \mathcal{C}$,
    algorithm ~\ref{alg:bqp_sim} simulates $C_{(\pi, g, \vec{s})}$
    on input $\ket{00 \dots 0}$ in time $\poly(\lvert C_{(\pi, g, \vec{s})} \rvert )$.
\end{corollary}
\begin{proof}
    By Proposition~\ref{hiddenshift_rewrite_existence},
    for each $C_{(\pi, g, \vec{s})} \in \mathcal{C}$ on $n$ qubits, we have that
    $
        \llbracket C_{(\pi, g, \vec{s})} \rrbracket \circ \ket{00 \dots 0}
        \stackrel{\ast}{\rewrite} \ket{\vec{s}}
    $
    which has no variables which is
    trivially $O(\log \lvert C_{(\pi, g, \vec{s})} \rvert )$.
    Thus by Lemma~\ref{bqp_sim_lemma}, algorithm~\ref{alg:bqp_sim}
    terminates in time $O(\lvert C_{(\pi, g, \vec{s})} \rvert^{2m+4})$.
    An $n$-fold application of algorithm~\ref{alg:bqp_sim} for each $i \in [n]$
    yields the shift $\vec{s}$, taking time
    \begin{align*}
        n \cdot O(\lvert C_{(\pi, g, \vec{s})} \rvert^{2m+4})
        = O(\lvert C_{(\pi, g, \vec{s})} \rvert^{2m+5})
    \end{align*}
\end{proof}

\begin{remark}
    It can be noted that in the case of hidden shift circuits,
    we can avoid the $n$-fold application of
    algorithm~\ref{alg:bqp_sim} by simply rewriting
    $\llbracket C \rrbracket \circ \ket{00 \dots 0}$ directly.
    Indeed, since hidden shift circuits are deterministic,
    we have that
    \begin{align*}
        eval \bigl( \llbracket C \rrbracket \circ \ket{00 \dots 0} \bigr)
        = \ket{s}
    \end{align*}
    hence in algorithm~\ref{alg:bqp_sim}, we can instead
    simply set $f \leftarrow g$ in the second line,
    so that the final evaluation yields $\ket{s}$ exactly,
    saving a factor of $n$ by applying the algorithm exactly once.
\end{remark}

\section{Conclusion}\label{sec:conclusion}
In this paper we have shown that a family of quantum circuits for Roetteler's
shifted bent function algorithm which requires non-Clifford resources and is
not trivially efficient to simulate can in fact be simulated
in polynomial time using path sums. We do so by giving a rewrite system on
path sums which is both restrictive enough to prove \emph{confluence}, while
also powerful enough to deterministically reduce implementations of
Roetteler's algorithm to the hidden shift in a small (linear) number of steps.
This rewrite system is applied within a simulation algorithm which
first simplifies a path sum to reduce the number of variables before explicitly 
evaluating the sum.
As simplifications of the path sums are closely related to tensor network contractions,
it stands to reason that our work can be seen as a type of strategy for tensor network
contraction. In this sense, a novel aspect of the work is that through confluence
we show not just the \emph{existence} of (an analogue of) a contraction order
which gives an efficient simulation, but that
\emph{every} order with the simple restriction yields an efficient simulation.
As our results apply to oracles implemented over a gate set which necessarily produce bent functions with multilinear polynomial representations having polynomially-many terms, the theoretical implications of this work appear to be limited. Moreover, as the query complexity separation between quantum and classical solutions to the shifted bent function problem do not necessarily translate to realistic quantum speedup, the utility of a polynomial-time classical simulation of such circuits is dubious. Instead, this work shows that this particular class of circuits is not a suitable benchmark for classical simulation algorithms, a point alluded to by \cite{koch_speedy_2023, codsi_cutting_2022}. More generally, our work suggests that the sum-over-paths approach may yield efficient simulations of certain classes of circuits. 

The effectiveness of rewriting in the context of general circuit simulation remains an interesting question. In the worst case, our rewrite system cannot simplify the path sum at all, leading to a very costly evaluation of the sum. It happens that nothing is lost in this case, as the resulting sum may then be decomposed into a sum of stabilizer sums, or more generally other polynomial-time simulable sums. It remains a question for future work to explore these decompositions in practice and benchmark them against other methods. More generally, it remains an open question to find other rewrite systems, ideally which satisfy strong guarantees on their time complexity to reach normal forms and determinism as the one we have developed here, which are effective in the context of circuit simulation. One potential avenue of simulation which was alluded to in \cite{amy_complete_2023} is to generalize the sum from variables to \emph{varieties} over $\mathbb{F}_2^k$, which allows the entire phase polynomial to be deterministically absorbed into an ideal of polynomial equations. In one sense this offloads the difficulty of simulation to the similarly difficult problem of computing Gr\"{o}bner bases and counting points in varieties, and so it remains a question for future work to explore this technique.

\section{Acknowledgements}

MA acknowledges support from the Canada Research Chair program, NSERC Discovery grant RGPIN-2022-03319, NSERC Alliance Quantum Software Consortium, and the Google Research Scholar program. MA and LSS thank the reviewers for their constructive comments and suggestions.
\bibliographystyle{abbrvurl}
\bibliography{references}

\appendix
\ifdefined \MAIN \else
    \documentclass{article}
    
    \moduleopen[Appendix]
\fi

\section{Multilinear Forms and Polynomials Over $\mathbb{F}_{2}$}~\label{appendix_multilinear_forms}
\ifdefined \MAIN \else
    \documentclass{article}
    
    \moduleopen
\fi
Here we review some facts which are relevant to multilinear polynomials.
We prove a fact that a degree-nonincreasing homomorphism between polynomial
rings over $\mathbb{F}_{2}$ induce a unique ring homomorphism between their
respective Boolean rings.

\begin{lemma}~\label{non_increasing_degree}
    Let $\phi : \mathbb{F}_{2}[V] \rightarrow \mathbb{F}_{2}[V]$
    be a homomorphism. Then $\phi$ is degree-nonincreasing on $\mathbb{F}_{2}[V]$
    iff it is degree-nonincreasing on linear forms in $\mathbb{F}_{2}[V]$.
\end{lemma}
\begin{proof}
    Forward direction is immediate.
    Conversely, we have that $\phi$ is degree-nonincreasing on all variables $v \in V$.
    Then if for $f = \sum_{I}c_{I}V^{I} \in \mathbb{F}_{2}[V]$,
    let $\deg(f) = \max \{deg(V_{I}) \suchthat c_{I} \neq 0\}$
    so that
    \begin{align*}
        \deg(\phi(f))
            &\leq \max \{deg(\phi (V_{I})) \suchthat  c_{I} \neq 0\} \\
            &\leq \max \{deg(V_{I}) \suchthat   c_{I} \neq 0\} \\
            &= \deg(f)
    \end{align*}
\end{proof}
\begin{lemma}
    Let $\phi : \mathbb{F}_{2}[V] \rightarrow \mathbb{F}_{2}[V]$
    be an isomorphism. Then $\phi $ preserves total degree on $\mathbb{F}_{2}[V]$
    iff it preserves total degree on linear forms in $\mathbb{F}_{2}[V]$.
\end{lemma}
\begin{proof}
    The forward direction is clear. Suppose that $\phi$ preserves
    degree on linear forms and assume for the moment that $\phi$ is linear
    in that for linear forms $y$, $\phi(y)$ is a linear form.
    Then $\phi$ can be viewed as an invertible linear transformation
    on $\mathbb{F}_{2}^{\oplus V}$. Then by Lemma
    ~\ref{non_increasing_degree}, $\phi$, and $\phi^{-1}$
    are degree-nonincreasing on $\mathbb{F}_{2}[V]$, so that $\phi$
    is degree preserving. The implication for general $\phi$
    follows by writing $\phi = \phi' \circ C$ for linear $\phi'$ and $C$
    which adds constants to each variable as appropriate.
\end{proof}
\begin{lemma}
    Let $\phi : \mathbb{F}_{2}[V] \rightarrow \mathbb{F}_{2}[W]$
    be a degree-preserving homomorphism, then
    $\phi$ induces a degree preserving homomorphism
    $\phi' : \mathbb{F}_{2}[V]/I_{V} \rightarrow \mathbb{F}_{2}[W]/I_{W}$
    and this association preserves composition.
\end{lemma}
\begin{proof}
    Let $\psi := \pi \circ \phi$ where
    $\pi: \mathbb{F}_{2}[W] \rightarrow \mathbb{F}_{2}[W]/I_{W}$
    is the projection map.
    Let $\alpha $ be an arbitrary element of $I_{V}$.
    Then $\alpha = \sum_{i}\alpha _{i}(v_{i}^{2} - v_{i})$,
    where $\alpha_{i} \in \mathbb{F}_{2}[V]$ and $v_{i} \in V$ are indeterminates.
    Since $\phi $ is degree preserving,
    $\phi(v_{i}) = \sum_{j}\beta_{j}^{i}w_{j} + c_{i}$, for some
    $\beta_{j}^{i}, c_{i} \in \mathbb{F}_{2}$.
    Noticing in characteristic $2$,
    \begin{align*}
        \phi(v_{i})^{2}
        &= \left( \sum_{j}\beta_{j}^{i}w_{j} + c_{i} \right)^{2} \\
        &= \sum_{j}\beta_{j}^{i}w_{j}^{2} + c_{i}
    \end{align*}
    We have
    \begin{align*}
        \phi(\alpha)
        &= \sum_{i}\phi(\alpha _{i})(\phi (v_{i})^{2} - \phi (v_{i})) \\
        &= \sum_{i}\phi(\alpha _{i})(
            \sum_{j}\beta_{j}^{i}w_{j}^{2} + c_{i}
            - \sum_{j}\beta_{j}^{i}w_{j} + c_{i}
        )\\
        &= \sum_{i}\phi (\alpha _{i})(
            \sum_{j}\beta_{j}^{i}(w_{j}^{2} - w_{j})
        )\\
        &= \sum_{i,j}\beta_{j}^{i}\phi(\alpha _{i})
            (w_{j}^{2} - w_{j})
         \in I_{W}
    \end{align*}
    Thus $\psi $ factors through $\mathbb{F}_{2}/I_{V}$, inducing $\phi'$
    as desired.
    \begin{align*}
        \begin{tikzcd}[wire types={n,n}, row sep=huge, column sep=huge, scale=2]
    \mathbb{F}_{2}[V] \arrow[r, "\phi" ] \arrow[d, "\pi "']
    & \mathbb{F}_{2}[W] \arrow[d, "\pi"] \\
    \mathbb{F}_{2}[V]/I_{V} \arrow[r, "\phi'"', dashed]
    & \mathbb{F}_{2}[W]/I_{W}
\end{tikzcd}

    \end{align*}
    By uniqueness of $\phi'$, it follows that compositions and inverses
    are preserved.
\end{proof}

It is a well-known fact that any polynomial ring over a unique factorization domain (UFD)
is a UFD.

\begin{lemma}[Pseudo Division in UFDs  - pg.54 of ~\cite{geddes_algorithms_1992}]
    ~\label{pseudo_division}
    Let $D[x]$ be a polynomial ring over a UFD $D$.
    For all $a(x), b(x) \in D[x]$ with $b(x) \neq 0$,
    there exist unique polynomials $q(x), r(x) \in D[x]$
    such that
    \begin{align*}
        \beta^{l} a(x) = b(x)q(x) + r(x)
    \end{align*}
    where $\deg(r(x)) < \deg(b(x))$, $\beta$ is the leading coefficient of $b(x)$
    and $l = \deg(a(x)) - \deg(b(x)) + 1$.
\end{lemma}

\newtheorem*{repeatlemma}{Lemma \ref{substitution_mirror}}
\begin{repeatlemma}
    Let $a, b \in \mathbb{F}_{2}[V]$ be polynomials
    such that $\deg(a) = \deg(b) = 1$.
    Then for $c \in \mathbb{F}_{2}$,
    $a[b \leftarrow 0] = c$ iff
    $a + b = c$.
\end{repeatlemma}
\let\repeatlemma\relax
\begin{proof}
    Let $a + b =c \in \mathbb{F}_{2}$,
    then $(a+b)[b \leftarrow 0] = a[b \leftarrow 0] = c$.
    Conversely let $a[b \leftarrow 0] = c \in \mathbb{F}_{2}$ and
    $x \in \Var(b)$ be the target variable
    of the substitution $[b \leftarrow 0]$.
    Then by pseudo-division, we have
    \begin{align*}
        a = qb + r
    \end{align*}
    where $x \notin \Var(r)$.
    Thus $a [b \leftarrow 0] = r [b \leftarrow 0] = r = c$.
    Since $\deg(a)=1$, it follows that $q = 1$, so
    \begin{align*}
        a = b + r = b +c
    \end{align*}
    and $a + b = c$.
\end{proof}

In particular $a [b \leftarrow 0] = c$ iff $b [a \leftarrow 0] = c$.


\moduleclose

\section{Correctness of Bravyi-Gosset Construction of Hidden Shift Circuits via Circuit Equalities}~\label{appendix_hiddenshift_circuit_proof}

We show that the implementation of Roetteler's algorithm with $\pi=\mathrm{id}$ in \cite{bravyi_improved_2016} can be reduced to the unique hidden shift using circuit equalities. As shown in the circuit-level proof below, the application of $X^{s}$ to apply a literal shift, as in \cite{bravyi_improved_2016} (note that $Z^s$ was equivalently used in their implementation), allows the two applications of $O_{f_0}:\ket{x}\mapsto (-1)^{f_0(x)}\ket{x}$ to cancel, at which point the rest follows from basic simplifications. 

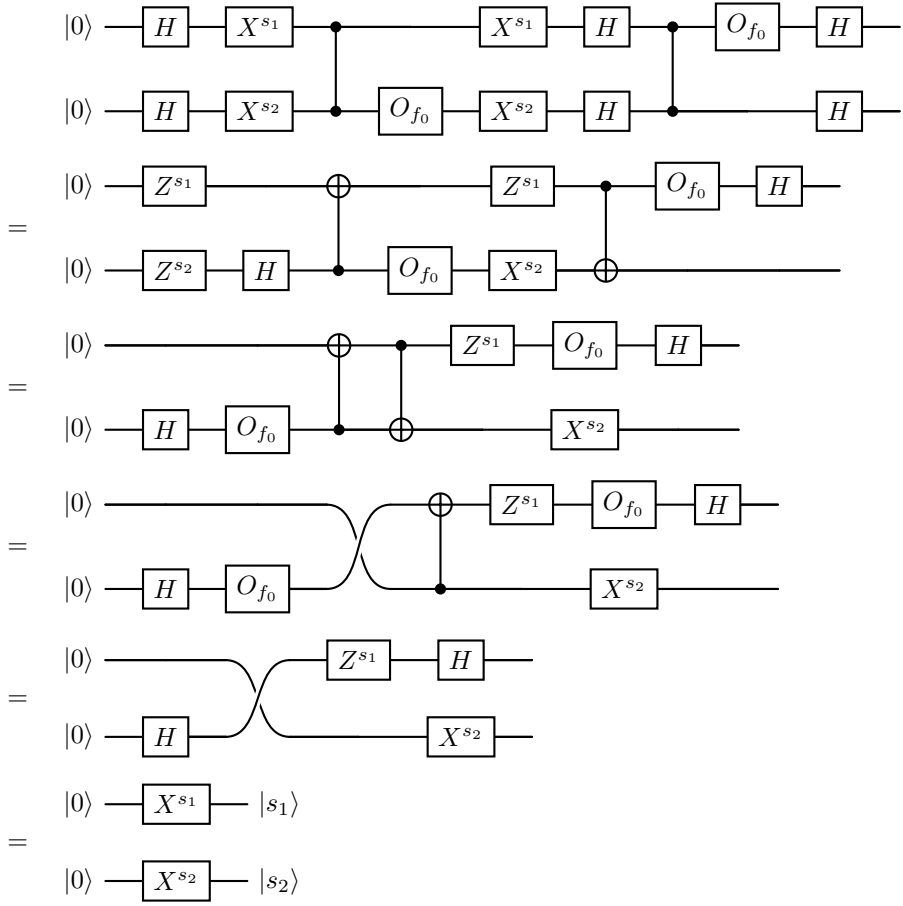
\begin{align*}
    &\begin{quantikz}[row sep=0.5cm, ampersand replacement=\&]
        \lstick{$\ket{0}$} \& \gate{H} \& \gate{X^{s_{1}}} \& \ctrl{1} \& \qw \& \gate{X^{s_{1}}} \& \gate{H}
        \& \ctrl{1} \& \gate{O_{f_0}} \& \gate{H} \& \qw\\
        \lstick{$\ket{0}$} \& \gate{H} \& \gate{X^{s_{2}}} \& \control{} \& \gate{O_{f_0}} \& \gate{X^{s_{2}}} \& \gate{H}
        \& \control{} \& \qw \& \gate{H} \& \qw
    \end{quantikz} \\
	\raisebox{0em}{$=$\quad}
    &\begin{quantikz}[row sep=0.5cm, ampersand replacement=\&]
        \lstick{$\ket{0}$} \& \gate{Z^{s_{1}}} \& \qw \& \targ \&  \qw \& \qw \& \gate{Z^{s_{1}}}
        \& \ctrl{1} \& \gate{O_{f_0}} \& \gate{H} \& \qw\\
        \lstick{$\ket{0}$} \& \gate{Z^{s_{2}}} \& \gate{H} \& \ctrl{-1} \& \gate{O_{f_0}} \& \gate{X^{s_{2}}}
        \& \targ \& \qw \& \qw \& \qw \& \qw
    \end{quantikz} \\
	\raisebox{0em}{$=$\quad}
    &\begin{quantikz}[row sep=0.5cm, ampersand replacement=\&]
        \lstick{$\ket{0}$} \& \qw \& \qw\& \targ \& \qw \& \ctrl{1} \& \gate{Z^{s_{1}}}
        \& \gate{O_{f_0}} \& \gate{H} \& \qw\\
        \lstick{$\ket{0}$} \& \gate{H} \& \gate{O_{f_0}} \& \ctrl{-1} \& \targ \& \qw
        \& \qw \& \gate{X^{s_{2}}} \& \qw \& \qw
    \end{quantikz} \\
	\raisebox{0em}{$=$\quad}
    &\begin{quantikz}[row sep=0.5cm, ampersand replacement=\&]
        \lstick{$\ket{0}$} \& \qw \& \qw\& \permute{2,1} \& \targ{} \& \gate{Z^{s_{1}}}
        \& \gate{O_{f_0}} \& \gate{H} \& \qw\\
        \lstick{$\ket{0}$} \& \gate{H} \& \gate{O_{f_0}} \& \qw \& \ctrl{-1} 
        \& \qw \& \gate{X^{s_{2}}} \& \qw \& \qw
    \end{quantikz} \\
	\raisebox{0em}{$=$\quad}
    &\begin{quantikz}[row sep=0.5cm, ampersand replacement=\&]
        \lstick{$\ket{0}$} \& \qw \& \permute{2,1} \& \gate{Z^{s_{1}}}
        \& \gate{H} \& \qw\\
        \lstick{$\ket{0}$} \& \gate{H} \& \qw
        \& \qw \& \gate{X^{s_{2}}} \& \qw
    \end{quantikz} \\
	\raisebox{0em}{$=$\quad}
    &\begin{quantikz}[row sep=0.5cm, ampersand replacement=\&]
        \lstick{$\ket{0}$} \& \gate{X^{s_{1}}}
        \& \rstick{$\ket{s_{1}}$} \\
        \lstick{$\ket{0}$} \& \gate{X^{s_{2}}} \& \rstick{$\ket{s_{2}}$}
    \end{quantikz}
\end{align*}

\section{Composition of Path Sums}~\label{appendix_composition}

We state useful lemmas relating compositions and tensor products
to the rewrites of their constituent parts. We also give two easy lemmas
pertaining to the time complexity of simplification and evaluation of path sums.

\begin{lemma}~\label{_compositional}
    Let $A, B\in \PS$ be path sums
    and suppose $A \rightarrow A'$.
    Then we have the following whenever compositions are well-defined.
    \begin{align*}
        A \circ B &\rightarrow A' \circ B \\
        A \otimes B &\rightarrow A' \otimes B \\
        B \circ A &\rightarrow B \circ A' \\
        B \otimes A &\rightarrow B \otimes A'
    \end{align*}
\end{lemma}
\begin{proof}
    We give the proof for the first statement
    when $A \rightarrow_{hh} D$. The rest follow similarly.
    Let $A$ have the general form of the LHS of the HH rule.
    \begin{align*}
        A &=  s_{A} \sum_{ V_{A} }
        (-1)^{x(y + Q) + R }
        \ket{O_{A}} \bra{I_{I}}
    \end{align*}
    Let $A: m \rightarrow n$, and $B: l \rightarrow m$.
    Then we have
    \begin{align*}
        A \circ B &=  \frac{s_{A}s_{B}}{2^{m}} \sum_{
            V_{A} \cup V_{B} \cup \{z_{1}, \dots, z_{m} \}
        }
        (-1)^{x(y + Q) + R + P_{B} + \sum z_{i}(I_{A,i} + O_{B,i})}
        \ket{O_{A}} \bra{I_{B}} \\
        &\rightarrow_{hh}  \frac{s_{A}s_{B}}{2^{m}} \sum_{
            (V_{A} \setminus \{x\})  \cup V_{B} \cup \{z_{1}, \dots, z_{m} \}
        }
        (-1)^{R[y \leftarrow Q] + P_{B} + \sum z_{i}(I_{A,i}[y \leftarrow Q] + O_{B,i})}
        \ket{O_{A}[y \leftarrow Q]} \bra{I_{B}} \\
        &= \frac{s_{D}s_{B}}{2^{m}} \sum_{
            V_{D} \cup V_{B} \cup \{z_{1}, \dots, z_{m} \}
        }
        (-1)^{P_{D}+ P_{B} + \sum z_{i}(I_{D,i} + O_{B,i})}
        \ket{O_{D}} \bra{I_{B}} \\
        &= D \circ B
    \end{align*}
    which follows from the fact that
    \begin{align*}
        D &=  s_{A} \sum_{ V_{A} \setminus \{x\}  }
        (-1)^{ R[y \leftarrow Q] }
        \ket{O_{A}} \bra{I_{I}} [y \leftarrow Q]
    \end{align*}
\end{proof}

\newtheorem*{repeatlemma}{Lemma \ref{compositional}}
\begin{repeatlemma}
    Let $A, B\in \PS$, be path sums such
    that $A \overset{\epsilon}{\rightarrow} A'$
    and $B \overset{\epsilon}{\rightarrow} B'$.
    Then
    \begin{align*}
        A \circ B &\rightarrow A' \circ B' \\
        A \otimes B &\rightarrow A' \otimes B'
    \end{align*}
    where $\overset{\epsilon }{ \rightarrow }$ is the reflexive
    closure of $ \rightarrow $, and $A \circ B$ is well-defined.
\end{repeatlemma}
\let\repeatlemma\relax
\begin{proof}
    Immediately follows from Lemma~\ref{_compositional}.
\end{proof}

\begin{lemma}
    Let $A \in \PS$ be a path sum with signature $n \rightarrow l$
    such that all constituent polynomials are bounded in degree by
    a constant $m \in \mathbb{N}$, then simplifying $A$
    to a normal form by $ \rightarrow $
    takes time
    \begin{align*}
        O((n+l+1) \lvert V_{A} \rvert^{m+2})
    \end{align*}
\end{lemma}
\begin{lemma}
    Let $A \in \PS$ be a path sum with signature $n \rightarrow l$
    such that all constituent polynomials are bounded in degree by
    a constant $m \in \mathbb{N}$, then $eval(A)$ is computed in time
    \begin{align*}
        O((n+l+1)\lvert V_{A} \rvert^{m+1}2^{\lvert V_{A}\rvert })
    \end{align*}
    and yields a linear operator with description size $O((n+l)2^{\lvert V_{A} \rvert})$.
\end{lemma}

\moduleclose

\end{document}